\documentclass[traditabstract]{aa}
\usepackage{textcomp}
\usepackage{graphicx,graphics}
\usepackage{savesym}
\usepackage{amsmath}
\usepackage{float}
\savesymbol{iint}
\usepackage{txfonts}
\restoresymbol{TXF}{iint}
\usepackage{time}
\usepackage{subfigure}
\usepackage{url}
\usepackage{hyperref}

\usepackage[usenames]{color}

\newcommand{\beq}{\begin{equation}}
\newcommand{\eeq}{\end{equation}}
\newcommand{\bea}{\begin{eqnarray}}
\newcommand{\eea}{\end{eqnarray}}
\begin{document}
\frenchspacing
\title{Brightness of the Sun's small scale magnetic field: proximity effects}
\titlerunning{Proximity effects}

\author{I.\ Thaler, H.C.\ Spruit}
\authorrunning{I.\ Thaler \& H.C.\ Spruit}


\institute{Max-Planck-Institut f\"{u}r Astrophysik,
  Karl-Schwarzschild-Str.\ 1,
  D-85748 Garching, Germany
}
\date{\today}

\abstract{The net effect of the small scale magnetic field on the Sun's  (bolometric) brightness is studied with realistic 3D MHD simulations. The direct effect of brightening within the magnetic field itself is consistent with measurements in high-resolution observations. The high `photometric accuracy' of the simulations, however, reveal compensating brightness effects that are hard to detect observationally. The influence of magnetic concentrations on the surrounding nonmagnetic convective flows (a `proximity effect') reduces the brightness by an amount  exceeding the brightening by the magnetic concentrations themselves. The net photospheric effect of the small scale field ($\approx-0.34\%$ at a mean flux density of 50 G) is thus negative. We conclude that the main contribution to the observed positive correlation between the magnetic field and total solar irradiance must be magnetic dissipation in layers around the temperature minimum and above (not included in the simulations). This agrees with existing inferences from observations.
}

\maketitle

%

\section{ Brightness variation of the Sun }

The brightness of the Sun (total solar irradiance at earth orbit, TSI) varies over its 11 yr magnetic cycle, by an amount of order 0.08\% (cf.\ Fr\"ohlich 2006). Such a variation is too small to have a direct effect on the Earth's climate, even if in addition to the 11 yr cyclic variation there were a systematic effect of this order sustained over centuries. Direct measurements of the Sun's global output (with space-based instruments) have only been available for the past ~30 years, however. A systematic trend over this period, if present, is below the variation between individual cycles. This has raised the question whether the cause of variation is understood well enough to extrapolate the effects detected so far to longer time scales in the past and into the future. 

The mechanisms by which magnetic fields influence the brightness of the solar surface have been known qualitatively for several decades (Spruit 1977, hereafter S77, Chiang \& Foukal 1983,  Spruit 1991). Detailed quantitative understanding has now become possible through advances in realistic 3D numerical MHD simulations of magnetic surface structures, such as sunspots and small sale magnetic fields structures (Carlsson et al.\ 2004, Keller et al.\ 2004, Steiner 2005, Pietarila Graham et al. 2009). 

Magnetic brightness changes of both signs are present (reduction in spots and pores, increase in small structures); their net effect on TSI cancels to about 80\%, with a small positive increase remaining. Since there is no theory for what determines the relative surface coverage of dark and bright magnetic structure, the current theoretical understanding of brightness mechanisms is still insufficient for extrapolations of the TSI record. 

Irrespective of this uncertainty, a good estimate of the brightness of the small scale magnetic field, as the main contributor to TSI variation, is called for.  In addition to the known mechanisms that cause small magnetic structures to increase the bolometric brightness of the solar surface, they also have effects on their nonmagnetic surroundings. These effects have not been studied much. We call them `proximity effects', and assess their possible importance for the net brightness variation of the Sun.

\section{Causes of brightness variation}
\subsection{Observations}
TSI correlates closely with identifiable magnetic structures on the surface. Composite long term records of observables, such as, areas covered by sunspots, active region fields and `active network', statistically explain over 90\% of the observed variance in TSI (e.g., Fr\"ohlich \& Lean 2004, Wenzler et al. 2006, Ball et al. 2012). This can be seen as evidence that the only detectable contribution to TSI variation are the local brightness contrasts of magnetic structures themselves, and that measurements of areas covered by these structures can thus be used as `proxies' for TSI variation. The statistical success of the correlation with surface structures, however, involves adjustment of free amplitude parameters for the proxies. Since it does not provide physical explanations of the effects either, it does not have much predictive power. Its use for extrapolations outside the time span of TSI measurements is therefore uncertain. 

A source of concern in interpreting the TSI record are possible longer term brightness variations of the quiet surface regions that are not covered by the proxies used. At the level of sensitivity needed, brightness measurement of magnetically quiet areas is not possible from the ground, owing to limited photometric accuracy. The space-based measurements of TSI are sensitive enough, but do not resolve any structures on the solar disk. Sufficient photometric accuracy has been achieved at $\sim 5\arcsec$ resolution by the balloon-borne Solar Bolometric Imager experiment (SBI, Foukal \& Bernasconi 2008). It has not found indications of any significant  brightness variations outside areas covered by magnetic fields. 

On smaller scales, arcseconds and less, brightness contributions from the immediate nonmagnetic environment of magnetic structures might be present that have escaped detection so far. Reports of changes in granulation related to magnetic activity, for example, have a long history. Macris et al. 1955, 1984) measured a decrease of granulation size with increasing solar activity. These results were not universally accepted, but more local changes in granulation in individual active regions are well documented. Granules appear smaller, with lower flow speeds (`anomalous' granulation, Macris 1979, Schmidt et al. 1988, Title et al. 1986, 1992, Kobel et al. 2012).  A plausible cause for these differences would be geometrical constraints imposed on the convective flow by nearby magnetic structures. Given these clearly detectable changes in the morphology of granulation, it would be somewhat surprising if the mean brightness of granulation were not affected as well, at some level (Spruit 1998). One might expect that the convective heat flux would be reduced by magnetic flow constraints, for example, and granules correspondingly darker . This would constitute a complication to be accounted for in the interpretation of TSI variations.  Direct detection of such effects at the required levels of a fraction of a percent is observationally quite challenging, but has recently become possible using space-based data (Kobel et al. 2012).

The direct effect of magnetic fields on the other hand, i.e., the local brightness of points on the surface where a magnetic signal is present, can be measured rather reliably. At the low levels of magnetic activity in quiet network the field consists of a small-scale mixture of opposite polarities. Identifying magnetic brightness changes in such regions therefore requires high spatial resolution. A conceptually straightforward measurement consists of adding up the contributions from magnetic bright points identifiable at the available spatial resolution (S\'anchez Almeida et al.\ 2010). This provides a lower limit since it underestimates the contribution from poorly resolved magnetic patches, and because these patches occur preferentially in regions that are darker than average: the intergranular lanes. 

Schnerr \& Spruit (2011) present a detailed study that takes these factors into account and does not rely on a feature identification process. Some of the magnetically quietest regions were studied, where the (unsigned) magnetic flux density is about 10 G. At this flux level, a net magnetic brightness increase (at 630nm) of $\approx 0.15$\% was found at disk center in data from the Swedish 1-m Solar Telescope (resolution 0\,\farcs 2), and $\approx 0.10$\% in lower resolution data from the Hinode satellite. 
These measurements only quantify the brightness in the magnetic field itself; brightness changes that might be present in the immediate nonmagnetic neighborhood of the magnetic structures are not included. The results therefore may represent {\em overestimates} of the magnetic brightening in quiet regions. 

\subsection{Simulations}
The small scale magnetic field is one of the obvious applications of realistic 3D MHD simulations, since they need to cover only relatively small areas of solar surface. The high spatial resolution required for convergence of the numerical simulations with observations has been achievable for almost a decade (Keller et al.\ 2004, Carlsson et al. 2004, Steiner 2005, de Pontieu et al.\ 2006). Detailed comparison with observations, e.g.,\ in de Pontieu et al.\ (2006), shows the remarkable degree of agreement that realistic simulations achieve in practically all aspects of the observations at the photospheric level. This gives confidence that more subtle questions like the proximity effect of anomalous granulation are within reach with current present computational resources.  The required statistical accuracy can be achieved by either simulating a sufficiently large area of magnetically affected granulation, or by following it over a sufficient number of granule life times.   

Afram et al (2011) studied the center-to-limb variation (CLV) of magnetic brightening of small scale magnetic structures with a realistic 3D MHD simulation. The result agrees with the qualitative predictions from the `flux tube' picture above and with the observed CLV of active regions. The net effect on the (bolometric) brightness of the solar surface was not highlighted explicitly in this work. As we find below, this is probably because it is a smaller effect, for which larger areas and/or a long integration times are required than for a test of the CLV. 

\section{Brightness of magnetic structures and their environment}

\subsection{Origin of magnetic brightenings}

\begin{figure}[t]
\includegraphics[width=0.5\textwidth]{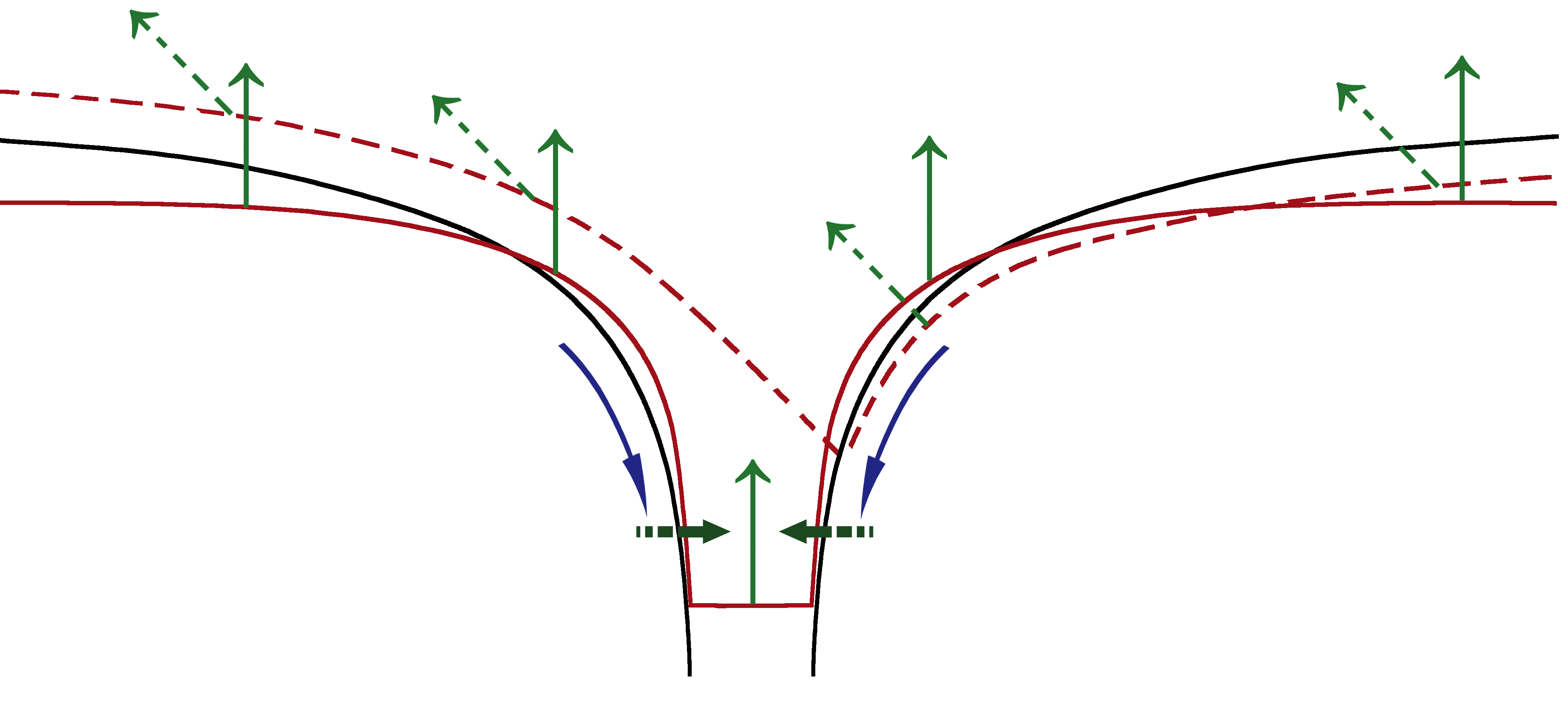}
\caption{Brightness changes in and around a magnetic `flux tube' (schematic). Black: boundary of the magnetic structure. Red: $\tau=1$ surfaces for viewing angles $\mu=1$ (solid) and $\mu=0.7$ (dashed). Green arrows: direction of the specific intensity from the $\tau=1$ surfaces for these angles.  At $\mu=0.7$ the `bright wall effect' is visible  towards the solar limb (right side of the figure), in the region where the  broken line lies below the solid line.  At the disk center (left) side, the interior of the tube is obscured by the wall of the tube. Lateral influx of heat into the flux tube (heavy broken arrows) cool the surroundings, causing enhanced downflow (blue arrows) around the structure.}
\label{sketch}
\end{figure} 

The reduced gas pressure in magnetic structures at the surface locally depresses the optical depth unity level (the `local photosphere'). This causes changes in surface brightness in two different ways. The reduced opacity causes a lateral influx of heat into the structure which starts to dominate the energy balance in sub-arcsecond size structures. Secondly, the geometric distortion of the local photospheric surface has an effect on the center-to-limb variation in surface brightness. When seen at an angle, i.e., at positions away from the center of the solar disk (Fig.\ \ref{sketch}), the walls of the structure are seen more nearly face-on, making them bright compared with the limb-darkened photosphere around it (the `bright wall effect', Spruit 1976, hereafter S76). This explains the increase of facular contrast observed towards the solar limb. It can be interpreted as a reduction of limb darkening caused by an increase of surface roughness. 

Because of the geometrical nature of the effects, the details of both also depend on the size of the structure.  This is illustrated in Fig.\ \ref{sketch}. A small structure (less than 0\,\farcs 5) at disk center is bright because of the sideways influx of radiation from the surrounding convection zone. Seen at an angle, the interior becomes obscured by the disk-center wall, while the limb side wall appears bright by its contrast with the limb-darkened surroundings. Fig.\ \ref{clv} shows the effect as seen in the numerical simulations reported below.

In a larger structure (pore, $\sim 1\arcsec$)  interference of the magnetic field with convective energy transport from below causes its center to be dark (as in sunspots).  Because of its greater width, its walls flare out more nearly horizontally over its surroundings. The opacity reduction effect increases the heat flux from this region. This brightness contribution turns out to be the dominant contributor to magnetic brightening away from disk center (Steiner 2005). It explains the rather large area affected ($\ga$ 0\,\farcs 5), compared with the small area expected from the height  ($\sim$ 150 km) of the walls of a narrow flux tube (see the sketch in Fig.\ \ref{sketch}). See the observations at {\tt\small  http://www.solarphysics.kva.se/gallery/movies/ 2004/gband\_10May2004\_AR\_limb.mpg} . Since smaller structures self-obscure away from disk center, the main brightness contribution towards the solar limb is actually from larger structures (pores) that are dark at disk center (see this in the images at {\tt\small http://www.solarphysics.kva.se/gallery/movies/2004/ gcont\_13May2004.avi} and S76 Fig.\,11). 

\begin{figure}
\includegraphics[width=0.5\textwidth]{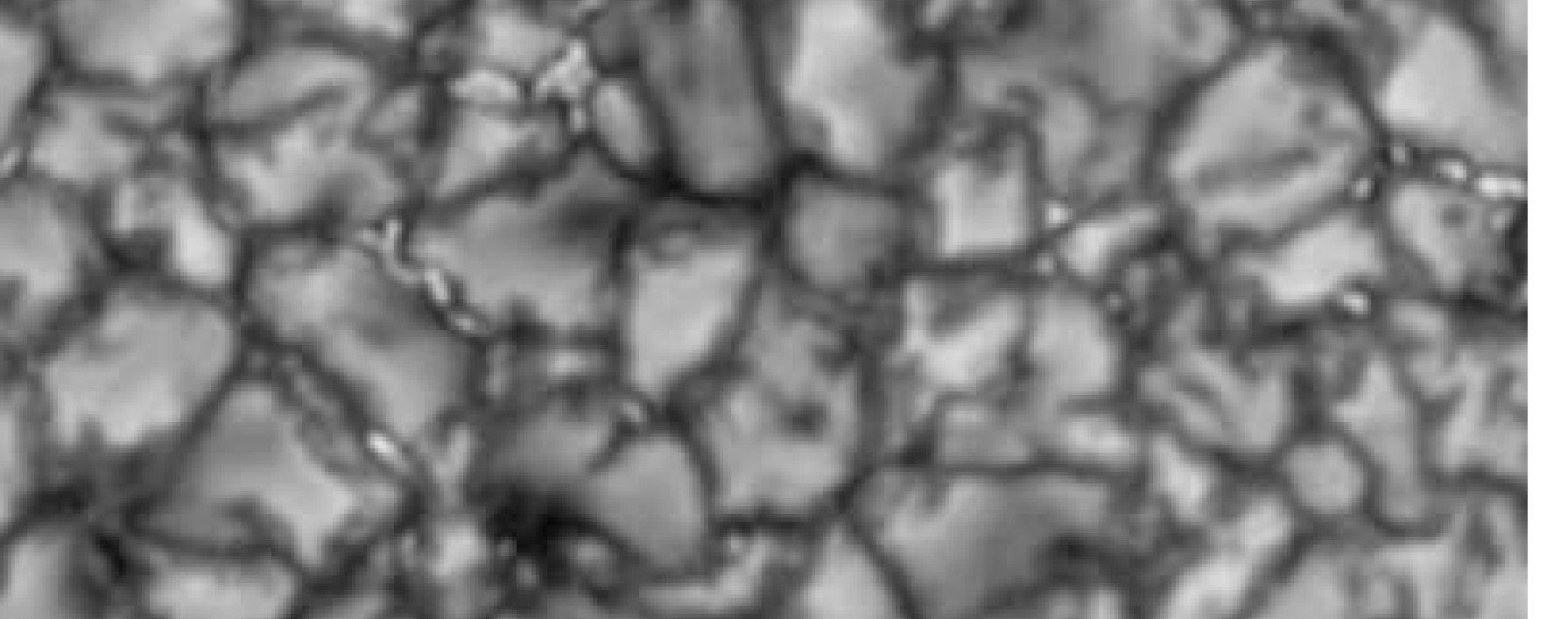}
\caption{Intensity image (630 nm continuum) of small scale magnetic fields at the disk center, as seen at $\approx0\, \farcs 15$ resolution (Swedish 1-m Solar Telescope). Image width is $20\arcsec$ (14.5 Mm). The magnetic brightenings between granules are surrounded by narrow dark rims: the `dark ring' effect (see \ Fig.\ \ref{sketch} for interpretation).}
\label{ring}
\end{figure} 

\subsection{Dark rings and dark lanes}
\label{rings}
The sideways radiative flux into a small magnetic structure derives in part from its immediate surroundings. This results (S76) in the presence of a `dark ring' in its immediate nonmagnetic environment. The theory predicted that this compensation is only partial, such that the small scale field acts as a net leak in the surface through which an excess heat flux escapes from the convection zone (S77).  The effect is predicted to be local, restricted to the immediate environment of the magnetic structures, not compensated by opposite brightness changes elsewhere on the solar surface: `what you see is what you get' (S77, Chiang \& Foukal 1983, S91, S97). The effect is clearly seen in high-resolution continuum images, see the example in  Fig.\,\ref{ring}. Quantitative assessment of the dark ring effect on the net brightness enhancement due to the magnetic field requires realistic numerical simulations. 

Measurement of the brightness effect of the small scale magnetic field is complicated by the fact that most of it is located in the intergranular lanes. Structures in intergranular lanes can add a positive brightness contribution even when they appear as darker than the mean nonmagnetic photosphere, at a given spatial resolution. Their effect on average brightness is determined by their  contrast relative to a comparable nonmagnetic location in the intergranular lanes. 

\begin{figure}
\includegraphics[width=0.5\textwidth]{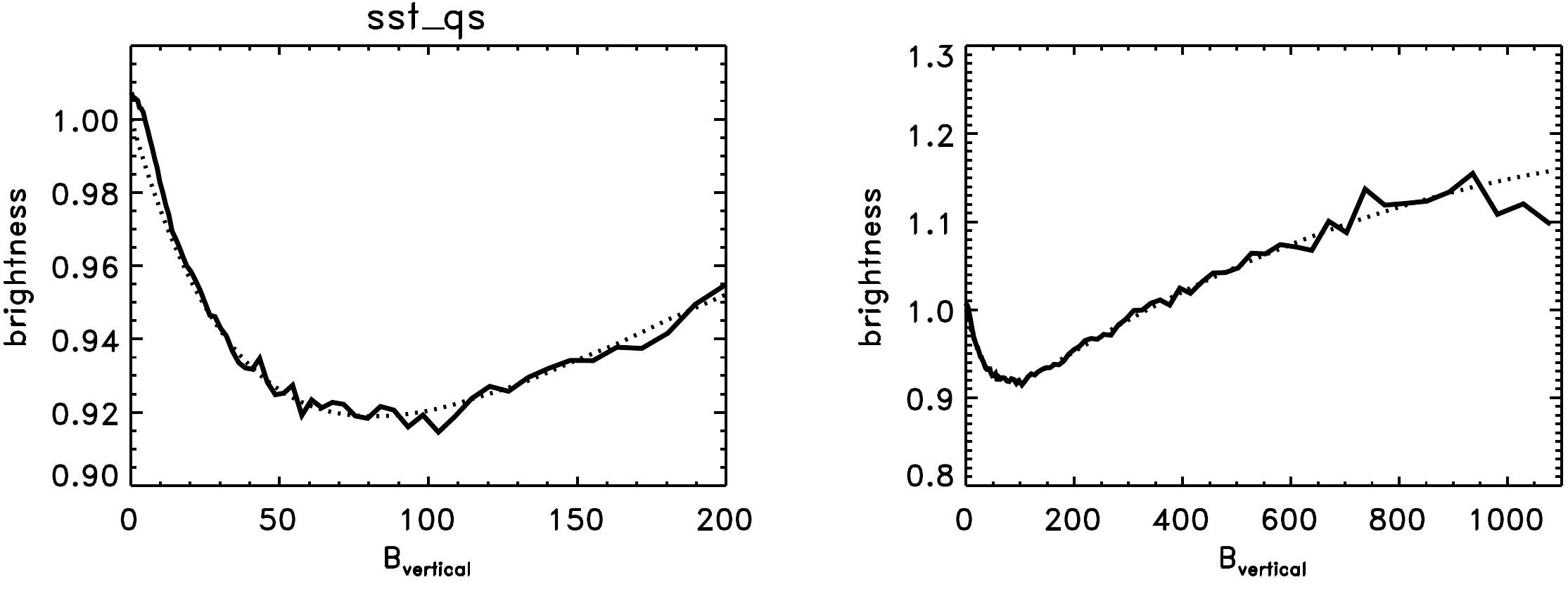}
\caption{Right panel, solid line: brightness as a function of flux density in a quiet region observed at 630 nm with the SST. Dotted: model fit as described in Schnerr and Spruit 2011. Left panel: same data on an expanded scale.}
\label{hookobs}
\end{figure}

The effect of this `dark lane bias' can be studied quantitatively by measuring the mean brightness of pixels of a given flux density  $\bar B$ (absolute value of the field strength averaged over the pixel) as a function of $\bar B$.  Fig.\ \ref{hookobs} shows this for observations made with the Swedish 1-m Solar Telescope (SST). The average (unsigned) flux density in the observation was 10 G. As expected, brightness increases with the amount of magnetic flux in the pixel, except  at flux densities below 100 G, where brightness first drops as a function of $\bar B$. The shape of the curve can be understood in terms of the dark lane bias: magnetic fields congregating in intergranular lanes (in particular at the vertices between several granules). 

At the very lowest flux densities (below $\approx 5$ G) the surface brightness is higher than the average by some 2\%. This is a consequence of the fact that magnetic fields preferentially populate intergranular lanes. The very lowest fields therefore avoid the lanes somewhat, causing a bias towards regions which are brighter than average. With increasing flux density, the bias shifts towards the intergranular lanes, causing the curve to drop below the average. The trend then reverses with increasing flux density, which selects pixels centered on the bright areas of larger, resolved, structures. [At even higher flux densities the curve dips down again, because in the field studied the largest field strengths occur in even larger, darker concentrations resembling pores, cf.\ Frazier (1971)]. 

From the model fit in Fig.\ \ref{hookobs} the brightening can be deduced, corrected for the dark lane bias. For this observation, this yields a net magnetic brightening of  the surface of  0.15\%, at a mean flux density of 11 G.

\section{Calculations}

\subsection{Numerical methods}
 For the numerical simulations we used the 3D
 magnetohydrodynamics code STAGGER, developed by
Galsgaard \& Nordlund (1996). The  code integrates the time-depended
 magnetohydrodynamics equations with a 6th order finite
 difference scheme using 5th order interpolations for the spatial
 derivatives. The  time evolution uses a 3rd order Runge-Kutta scheme. For every time step the radiative transfer equation is solved at every grid point in the 3D box assuming local thermodynamic equilibrium. This is done by using a Feautrier-like scheme along the rays with two $\mu$ angles plus the vertical and four $\phi$ angles horizontally, which adds up to nine angles in total. The wavelength dependence of the absorption coefficient is represented by four opacity bins. A more detailed description on the opacity binning scheme used is given in Collet et al. (2011). Further details about the code performance can be found in Beeck et al. (2012).

\subsection{Setup}

The horizontal boundaries are periodic while the vertical boundaries are open and transmitting. The entropy of the inflowing material at the bottom boundary is fixed and the same in all three simulations. The magnetic field is kept vertical at the bottom, allowing field lines to move horizontally there. As a magnetic boundary condition a potential field is implemented at the top boundary. 

As a reference for measuring the magnetic brightness effects a purely hydrodynamic simulation run was started from an already thermally relaxed snapshot. This simulation was run for 960 minutes of solar time. 
The initial condition for the magnetic simulations 
 is again a thermally relaxed snapshot. It is taken from the nonmagnetic reference run, but with a uniform vertical magnetic field added.  Since the Lorentz force in this field vanishes, adding such a field is consistent with the hydrodynamic part of the initial conditions. Due to the periodic boundary conditions, the horizontal mean of the vertical field component stays constant in time at all depths. Two magnetic simulations were done, with mean fields of 50 and 100 G. The intitial conditions for the two cases were taken from different, but statistically equivalent, points in the nonmagnetic reference run. Except for the initial conditions as described, the parameters controlling input physics and numerical setup are identical in all three runs.

For the main results reported below we used a simulation box of  horizontal extent 18 Mm x 18 Mm with a resolution of 25 km and a vertical extent of 3.15 Mm 
extending 465 km above the photosphere and 2.7 Mm into the convection zone. The grid is nonequidistant in the  vertical direction, with grid spacing varying between a minimum of 7 km near the photosphere  to 32 km at the lower boundary.  A few short runs at higher resolution  were done in connection with the physical interpretation of the brightness effects, see section \ref{measuprox}.

Since the time step in the magnetic simulations is determined by the Alfv\'en speed in the atmosphere, computational expense increases roughly with the initial field strength. The length of the 50 G run is 300 minutes of solar time, the 100 G simulation 120 min. These durations were needed to average out the realization noise to an acceptable statistical significance level.

\section{Results}

\begin{figure*}[t]
\center{\includegraphics[width=0.8\textwidth]{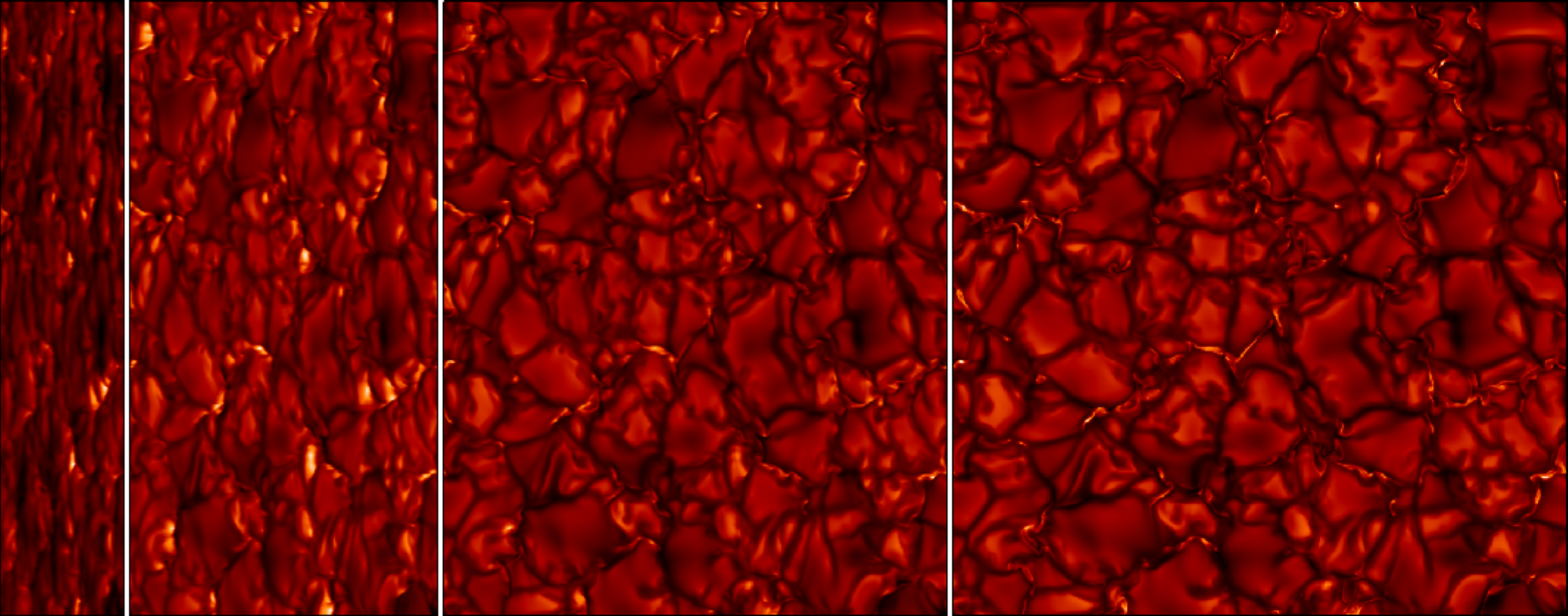}}
\caption{Snapshot images from the 100 G simulation, showing  the emergent specific intensity  in the  continuum at 630 nm, at viewing angles (left to right) $\mu=0.2,~0.5,~0.82$, and $1.0$. Height of the image is 18 Mm. (Center-to-limb variation of average brightness has been removed).}
\label{compo}
\end{figure*} 

\begin{figure}[b]
\includegraphics[width=0.5\textwidth]{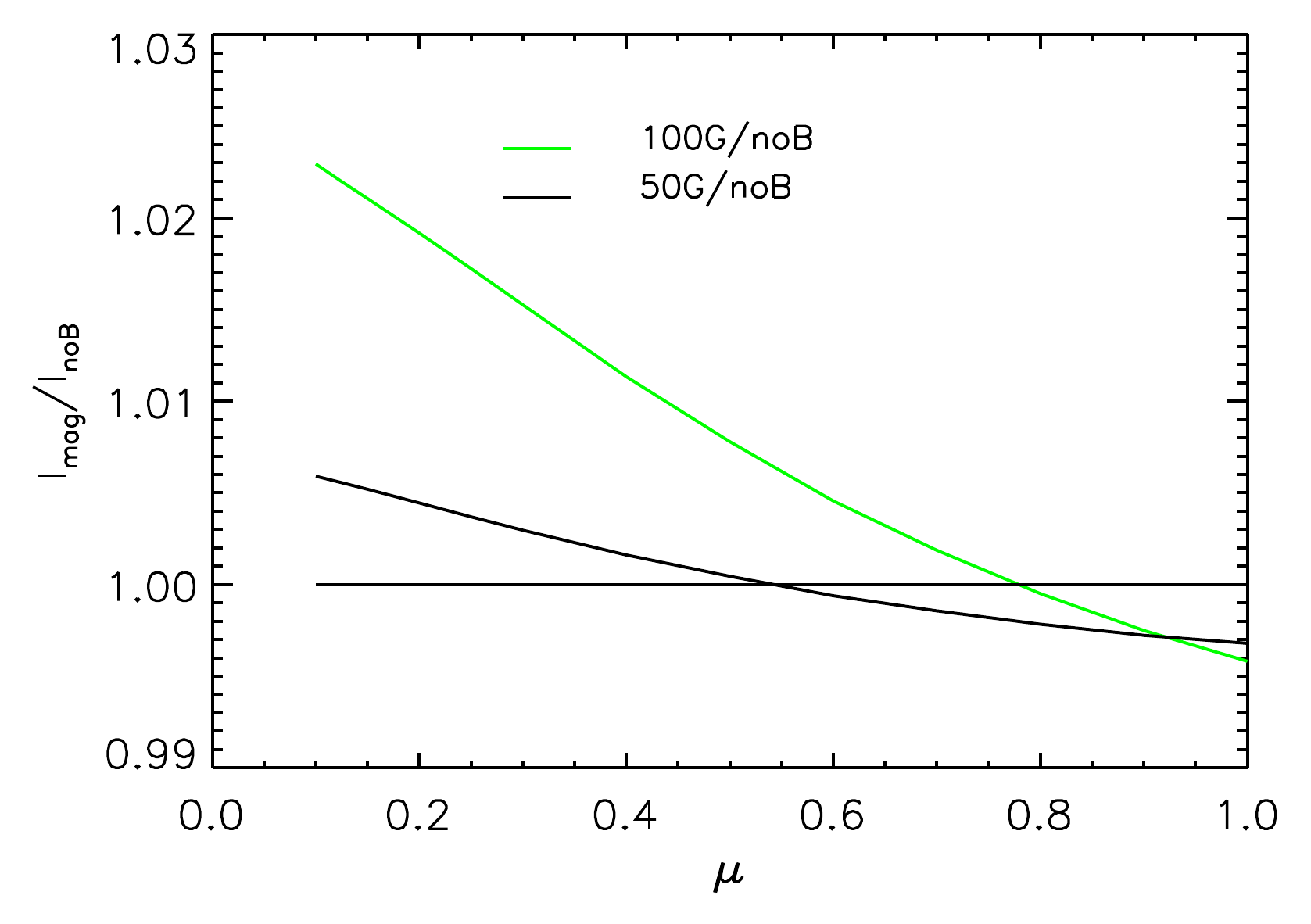}
\caption{Time and area-averaged center-to-limb variation of brightness in 630 nm continuum, relative to the nonmagnetic simulation.}
\label{clv}
\end{figure} 

\begin{figure}[t]
\hfil\includegraphics[width=1\hsize]{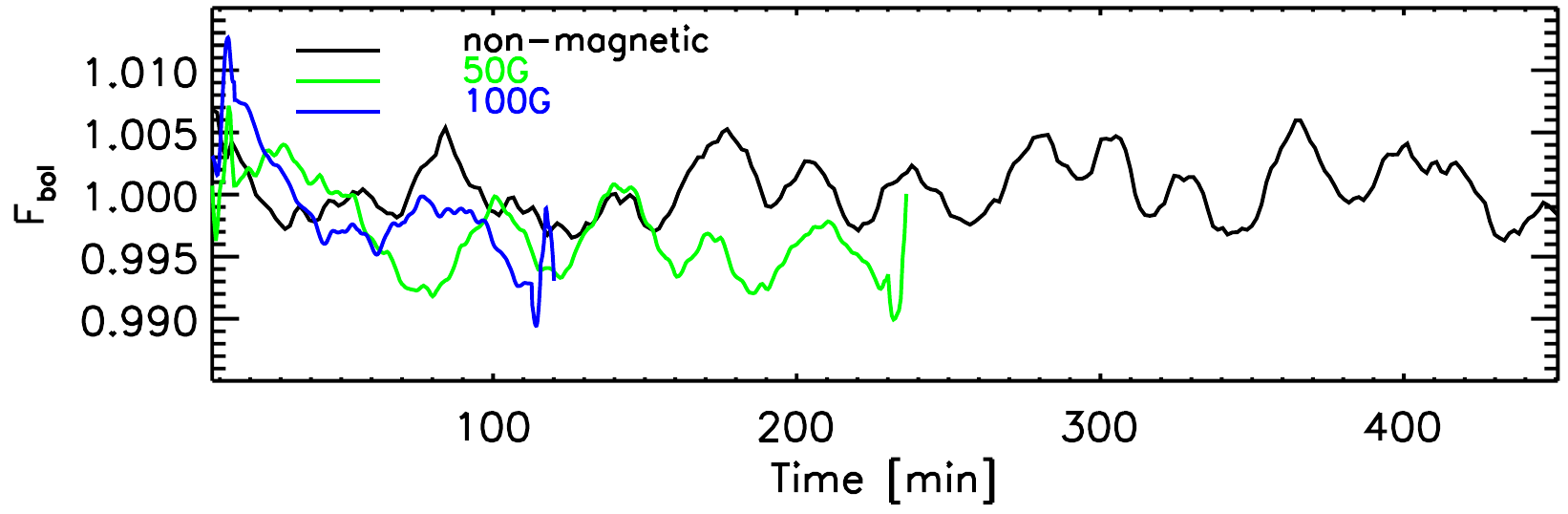}\hfil
\caption{Time evolution of the bolometric flux, averaged over the $18\time 18$ Mm$^2$ simulation area, running means over 15 min. 100 G simulation (blue), 50 G simulation (green) and the nonmagnetic simulation (black). Fluxes normalized to the mean of the nonmagnetic run.  }
\label{time_series}
\end{figure}

\begin{figure}
\includegraphics[width=0.25\textwidth]{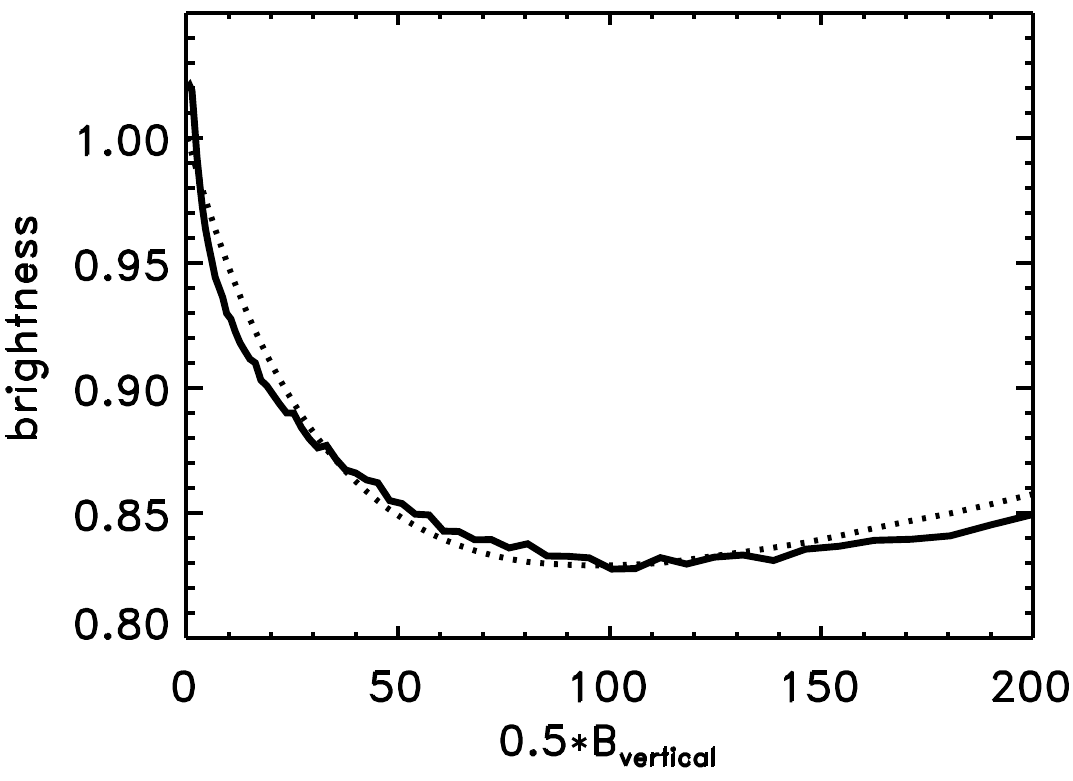}\includegraphics[width=0.236\textwidth]{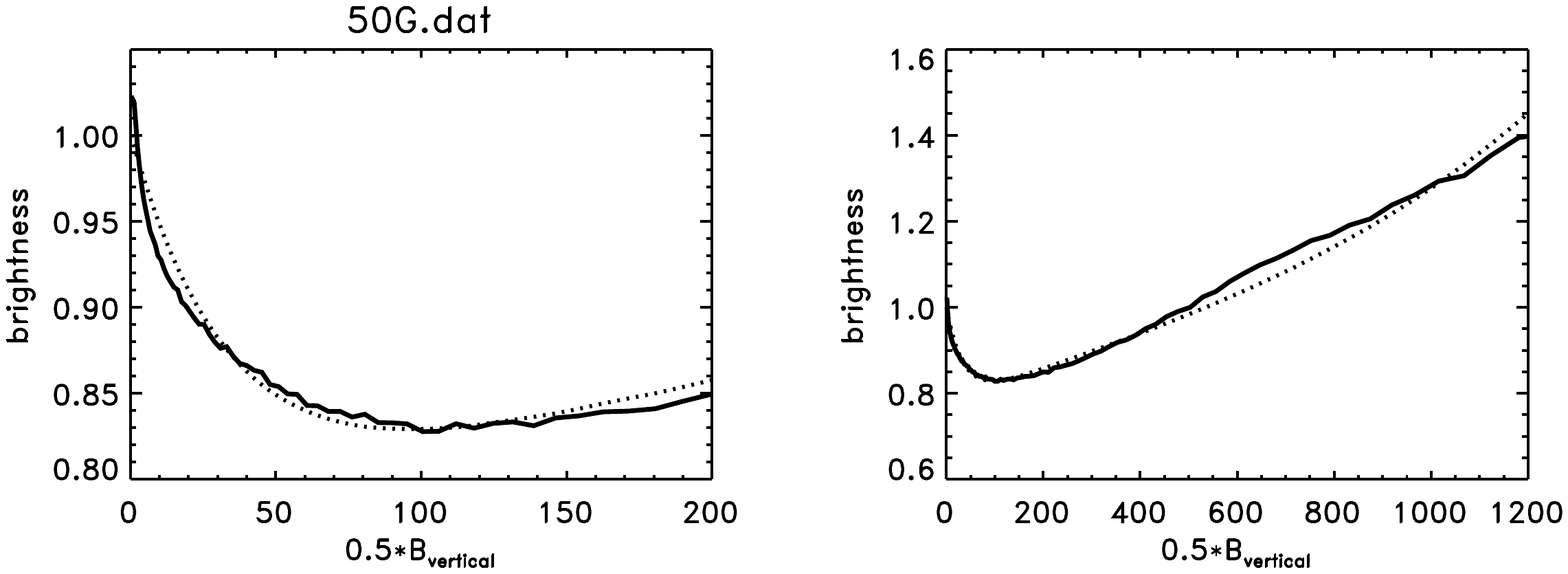}
\caption{As in Fig.\ \ref{hookobs}, for the 50 G simulation. The field strength scale shown corrects approximately for the difference in field strength at the nominal photosphere in the simulations (around 500 nm continuum optical depth unity) and the effective measurement level in the 630.25 nm line used in the observations.}
\label{hooksim}
\end{figure} 

Fig.\ \ref{compo}, shows a snapshot of the 100 G mean field simulation at 4 viewing angles. The smallest scales in the magnetic structures disappear from view already around $\mu=0.82$ ($35^\circ$ from disk center), resulting in a somewhat fuzzier impression. The dependence on $\mu$ shows the characteristic 'bright wall effect' that becomes conspicuous at $\mu\la0.7$. Fig.\ \ref{clv} shows the center-to-limb variation of the surface-averaged continuum brightness relative to that of the nonmagnetic simulation. At least 12 snapshots taken 10 min apart of each other were used for each simulation.

The positive contrast towards the limb is as expected from the bright wall effect. At disk center, however, the contrast is negative, on average. Though statistical fluctuations in granulation and the 5-minute oscillation can occasionally yield a positive mean brightness at disk center, the time-average is stably negative around disk center. This disagrees with the results reported by Afram et al. 2011 (their Fig. 8), which imply a net positive brightening at disk center by as much as 1\% even for 50 G mean field. This is probably in conflict with observations.

\begin{figure}
\center{\includegraphics[width=0.7\hsize]{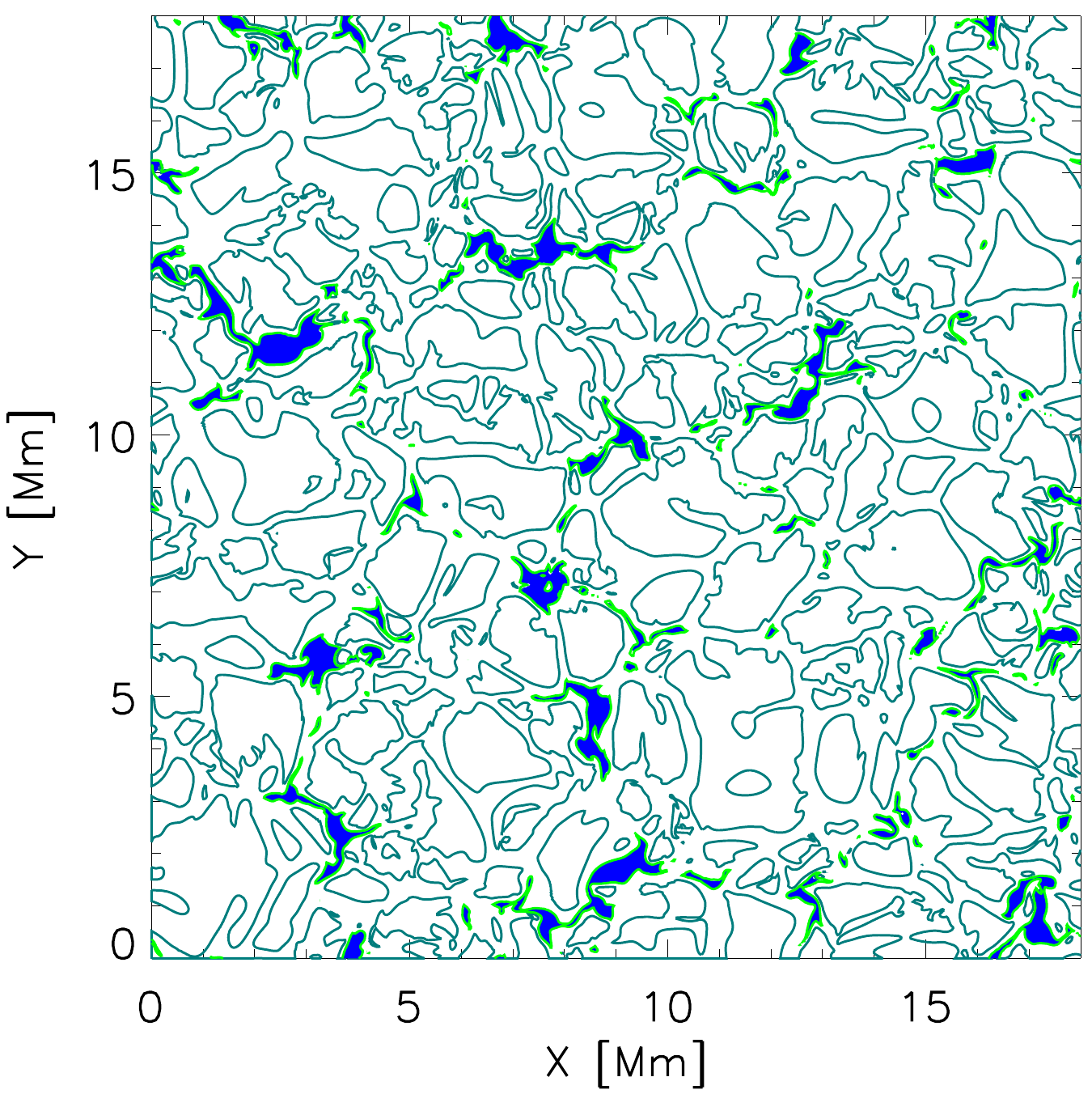}}
\center{\includegraphics[width=0.7\hsize]{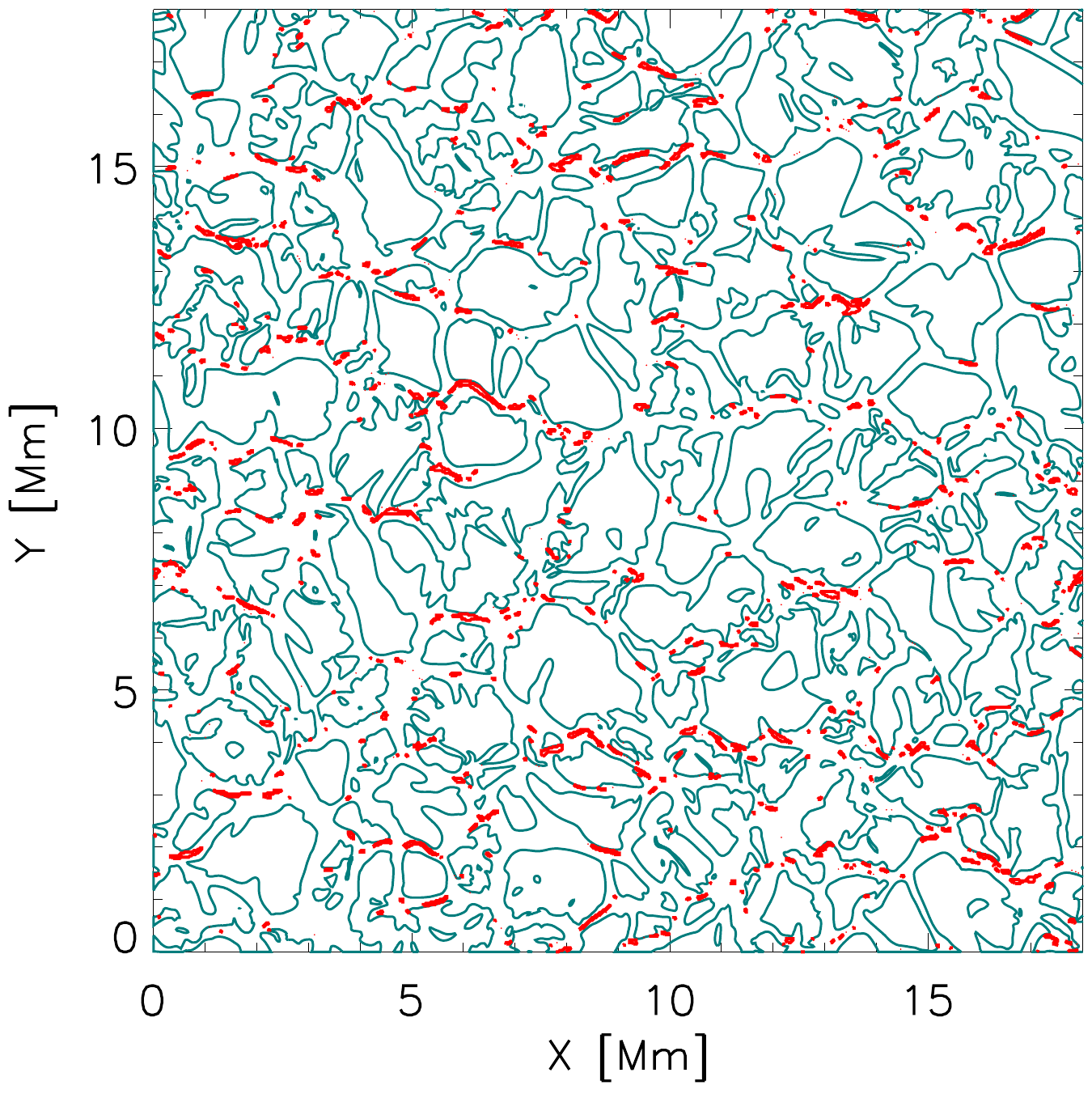}}
\caption{Magnetic and stagnation point selection. The gray contours  (at $v_z=0$) enclose the upflows. Top panel:  snapshot from the 100 G simulation. Green contours are at a vertical field strength of 1kG, the blue contours enclose the centering pixels of our magnetic structures. Bottom panel:  a snapshot from the nonmagnetic simulation. Pixels (red) show the regions of flow convergence  used as reference ($-1/{\rm div}\, {\bf v}_{\rm h}  < 26$  s).}
\label{area_sel}
\end{figure}

The  time evolution of the bolometric surface flux  of the three simulation runs is shown in Figure \ref{time_series}. Initially, Lorentz forces are absent, the magnetic field has no effect on the flow, and the flux level is unaffected. With time, the field gets concentrated into the intergranular lanes,  and the Lorentz forces start having an effect on the flow. After a few granule turnover times (20 min, say) the magnetic field and the flow pattern have settled to a state which accommodates the magnetic constraints. This explains the drop in brightness of the magnetic simulations over the first couple of turnover times.

The measured difference between the 50 G simulation and the nonmagnetic simulation is $\Delta F_{50\rm G-noB}= (-0.34 \pm 0.07)\%$, and $\Delta F_{100\rm G-noB}= (-0.27 \pm 0.09)\%$ in the 100 G  simulation. The error bars in these numbers were computed by assuming that the average life time of a granule is 10 min, and using this to evaluate the number of independent points in the curves. To find out how this darkening comes about a closer look at  the results is needed.

\subsection{Sources of darkening}
The darkening effect seen in Fig.\ \ref{time_series} is the opposite of the  expected magnetic brightening effect. This shows that the simulations include effects that have not been detected so far in the observations (see however Kobel et al.\ 2012). As a check on the reliability of the  calculations, we invoke the procedure that was used  in Schnerr and Spruit (2011) for measuring magnetic brightening in weak fields from high-resolution observations (see sect.\ \ref{rings}). The result of applying this same procedure to the simulations is shown in Fig.\ \ref{hooksim}. The model fit (dotted line, cf.\ sec.\ \ref{rings}) predicts a net magnetic brightening of 0.7\% for  the mean flux density of 50 G of the simulation. Assuming that the effect is proportional to the filling factor of the magnetic concentrations, this number translates  to 0.12\% for a flux density of 10 G. This agrees well with the number found for the observations in  Schnerr and Spruit (2011), where  the procedure yielded a brightening of 0.15\% in an area with a mean unsigned flux density  of 11 G.

This confirms that methods which focus on magnetic brightenings, whether in simulations or real data, miss the more subtle darkening effect. It is of lower amplitude, but extends over a larger area.
The Schnerr \& Spruit procedure, for example, corrects for the dark lane bias (thereby increasing the inferred magnetic brightening effect), but does not account for the proximity effect on the surrounding granulation flow (which reduce it).

\subsection{Measuring proximity effects}
\label{measuprox}
Inspection of an image gives a qualitative impression of the amplitude and spatial extent of  darkening near magnetic concentrations.
To quantify these proximity effects we need a way to average out the individual random brightness variations near the concentrations. We do this first by simply superposing and averaging a large number of subareas centered on magnetic pixels from the time series of images. Call this the image superposition method. Our selection criteria for centering are a vertical magnetic field strength above 1 kG and a ratio between the horizontal field strength to the total field strength below 0.5, as the magnetic field in the center of the magnetic structures is nearly vertical. Magnetic concentrations consist of clumps of neighboring pixels satisfying these criteria. Since the darkening effects investigated plausibly scale approximately linearly with the amount of magnetic flux, this selection adds the correct weight to the individual pixels in a magnetic concentration.  No attempt is made to determine the `centers' of the clumps, which in theory might yield a better measure of distance from a concentration. In practice, this would not help much since the clumps are actually narrow filaments. Unavoidably, their random orientation causes a certain horizontal averaging of the resulting image. The fine dark edges around the crinkles seen in Figs.\ \ref{ring}, \ref{compo} are completely smeared out. 

\begin{figure*}[t]
\includegraphics[height=0.255\textwidth]{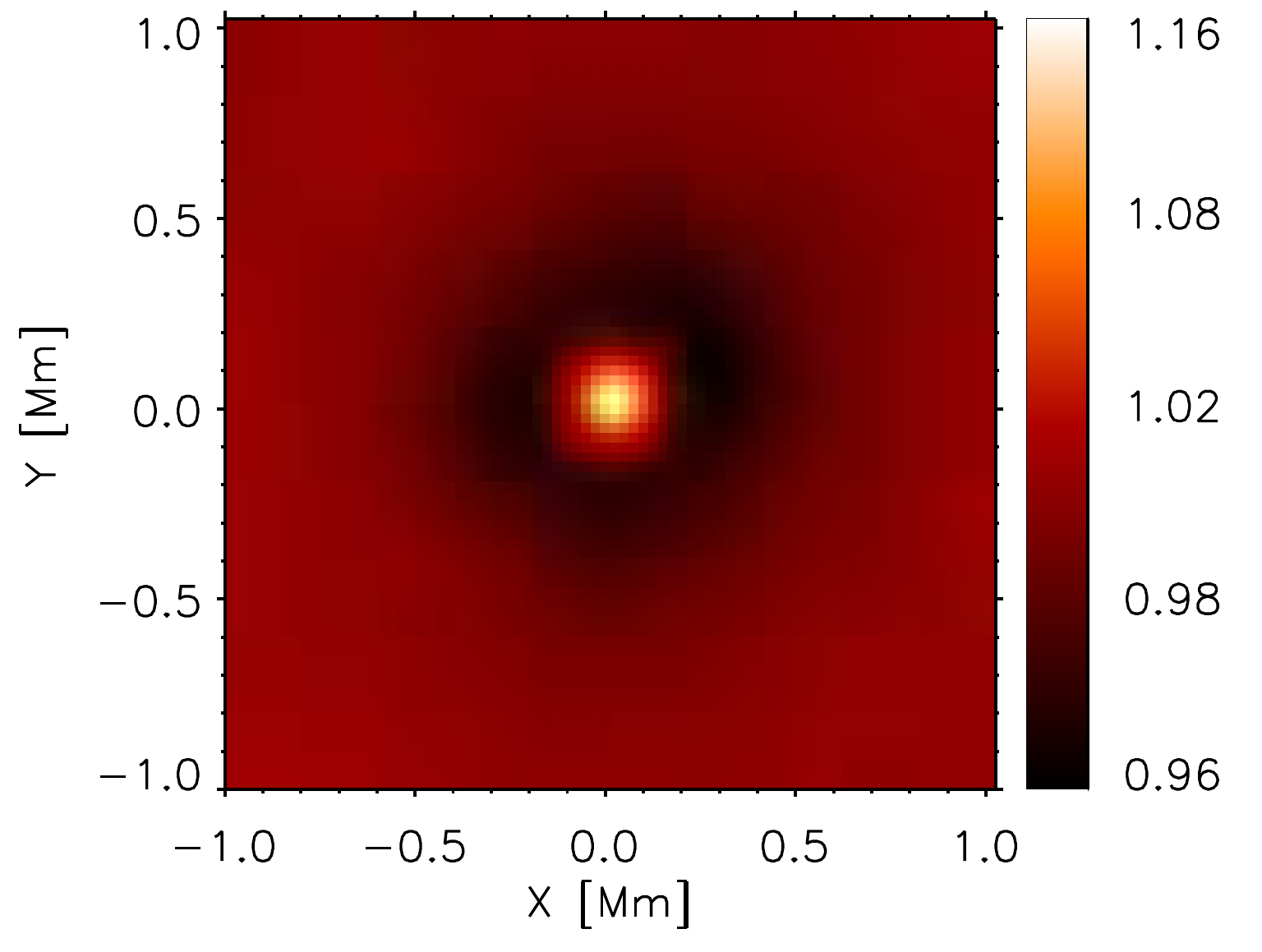}
\includegraphics[height=0.255\textwidth]{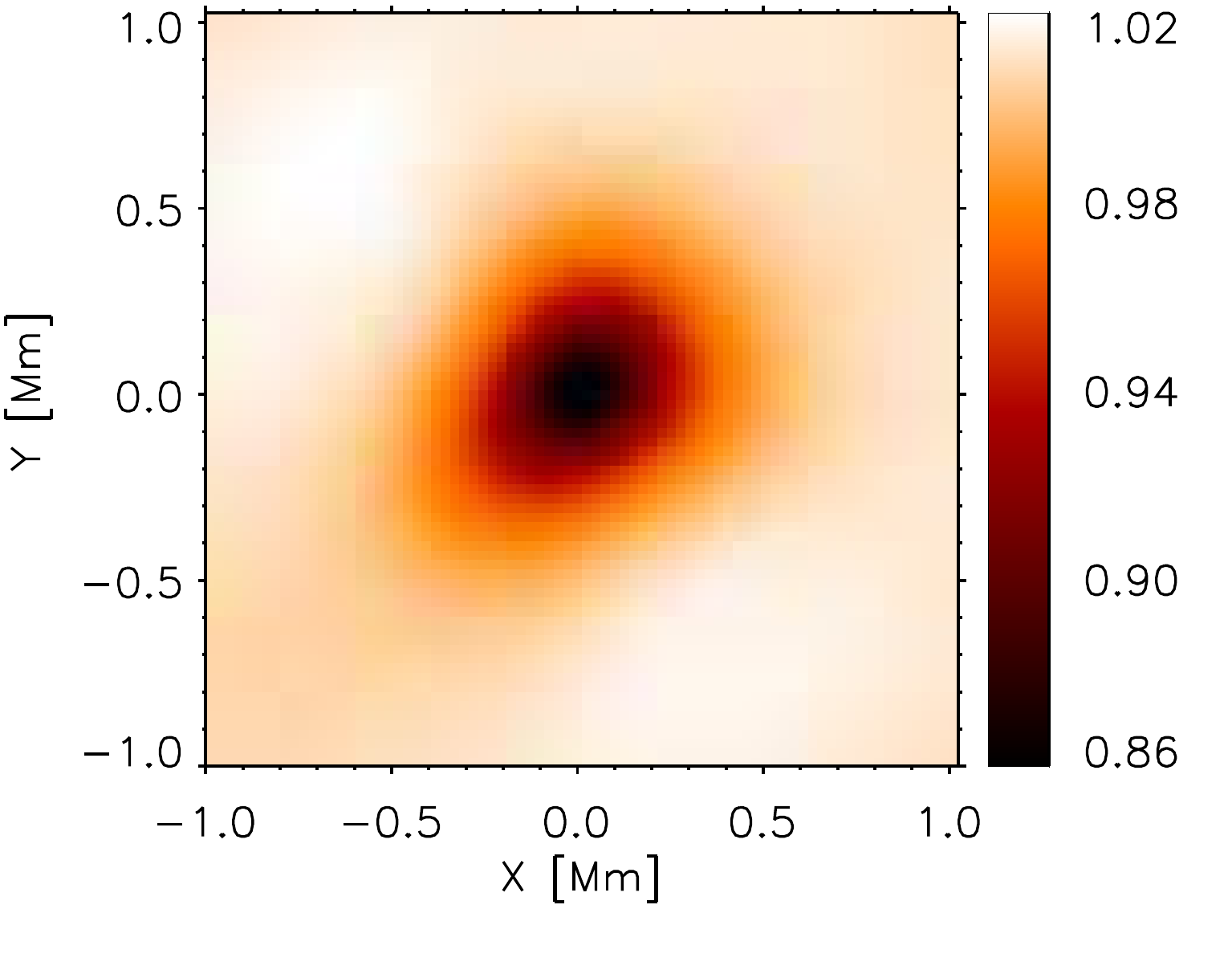}
\includegraphics[height=0.255\textwidth]{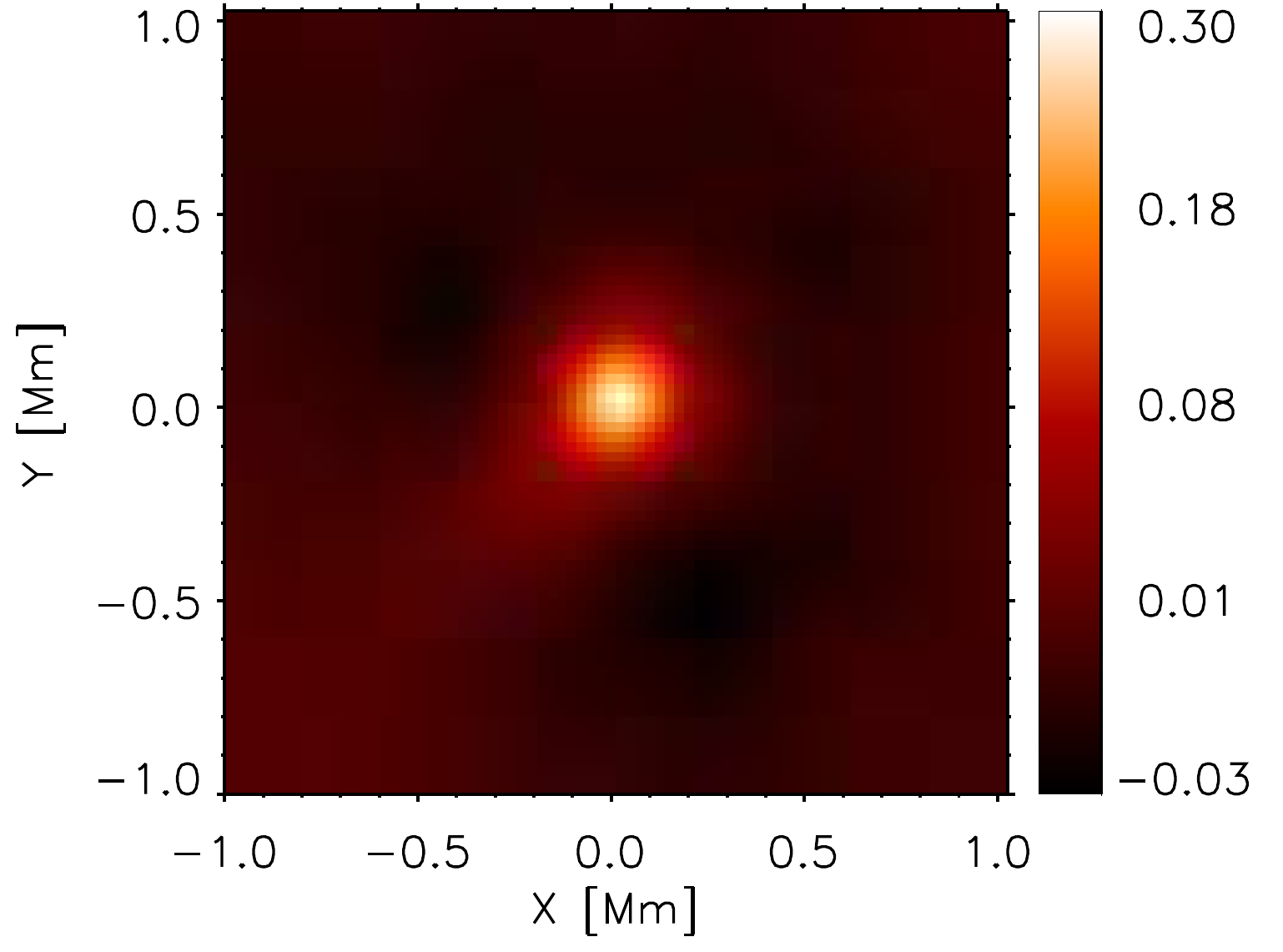}
\caption{Average brightness variation near magnetic structures in the 50 G simulation (left). Middle panel shows the equivalent in control areas (the stagnation points of the convection flow as seen in the nonmagnetic simulation). The difference (right) shows the net brightness effect of the magnetic concentrations.}\label{ringims}
\end{figure*}

Figure \ref{area_sel} (top panel) shows an example snapshot with the magnetic areas selected. The contours of the upflows are shown in gray. The green lines indicate areas with a field strength higher than 1 kG, while the blue colored regions show the pixels belonging to magnetic concentrations according to our selection criteria. 
 Figure \ref{ringims} (left panel) shows the average of a 2000 km wide area centered on the selected magnetic pixels, from the 50 G simulation. As expected, there is a conspicuous positive brightness contribution in the center of the magnetic structure, for 50 G simulation reaching up to  a factor of 1.16 of the average bolometric flux value. 

\begin{figure}[t]
\center{\includegraphics[width=0.9\hsize]{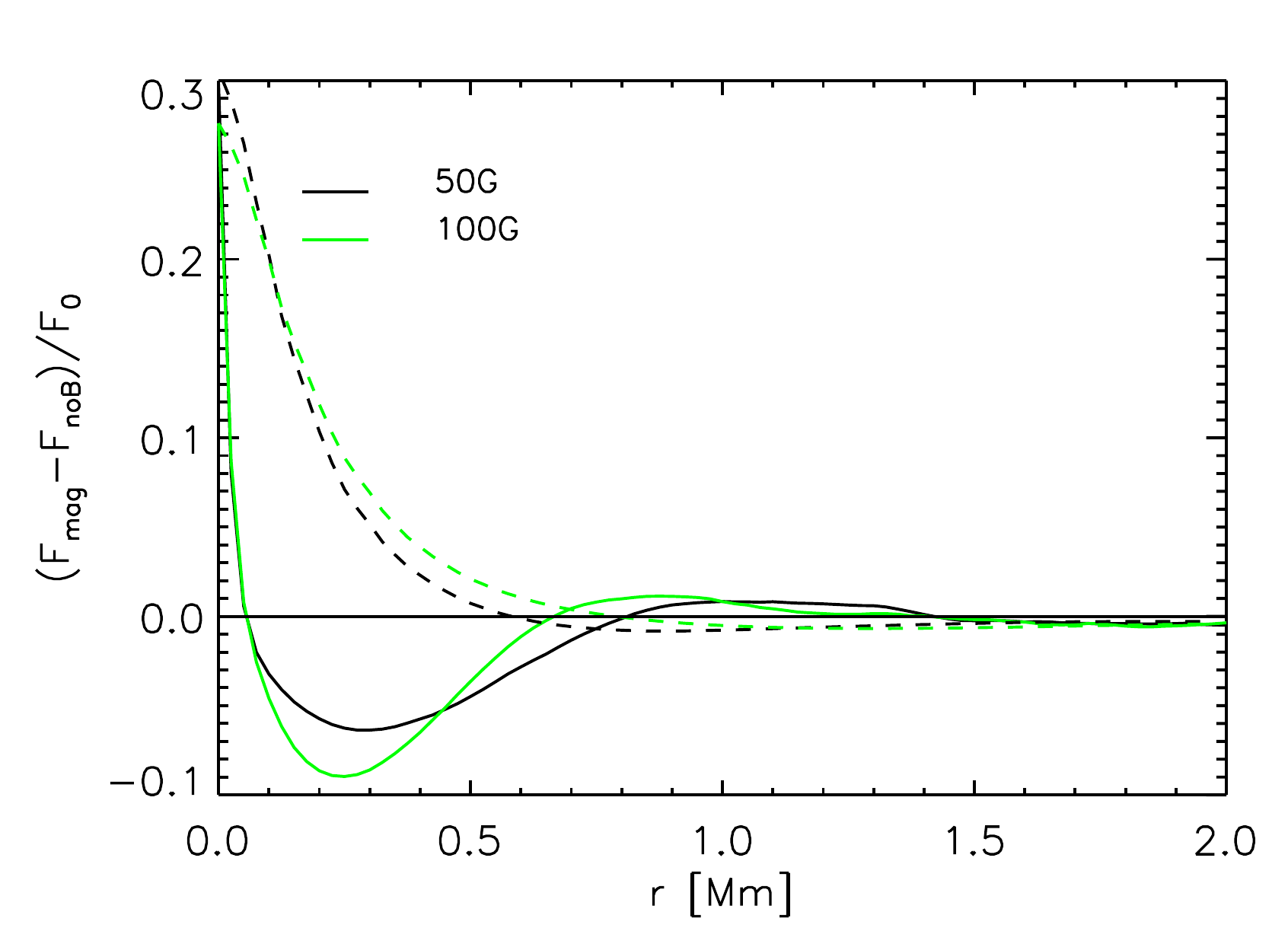}}
\hfill\includegraphics[width=0.9\hsize]{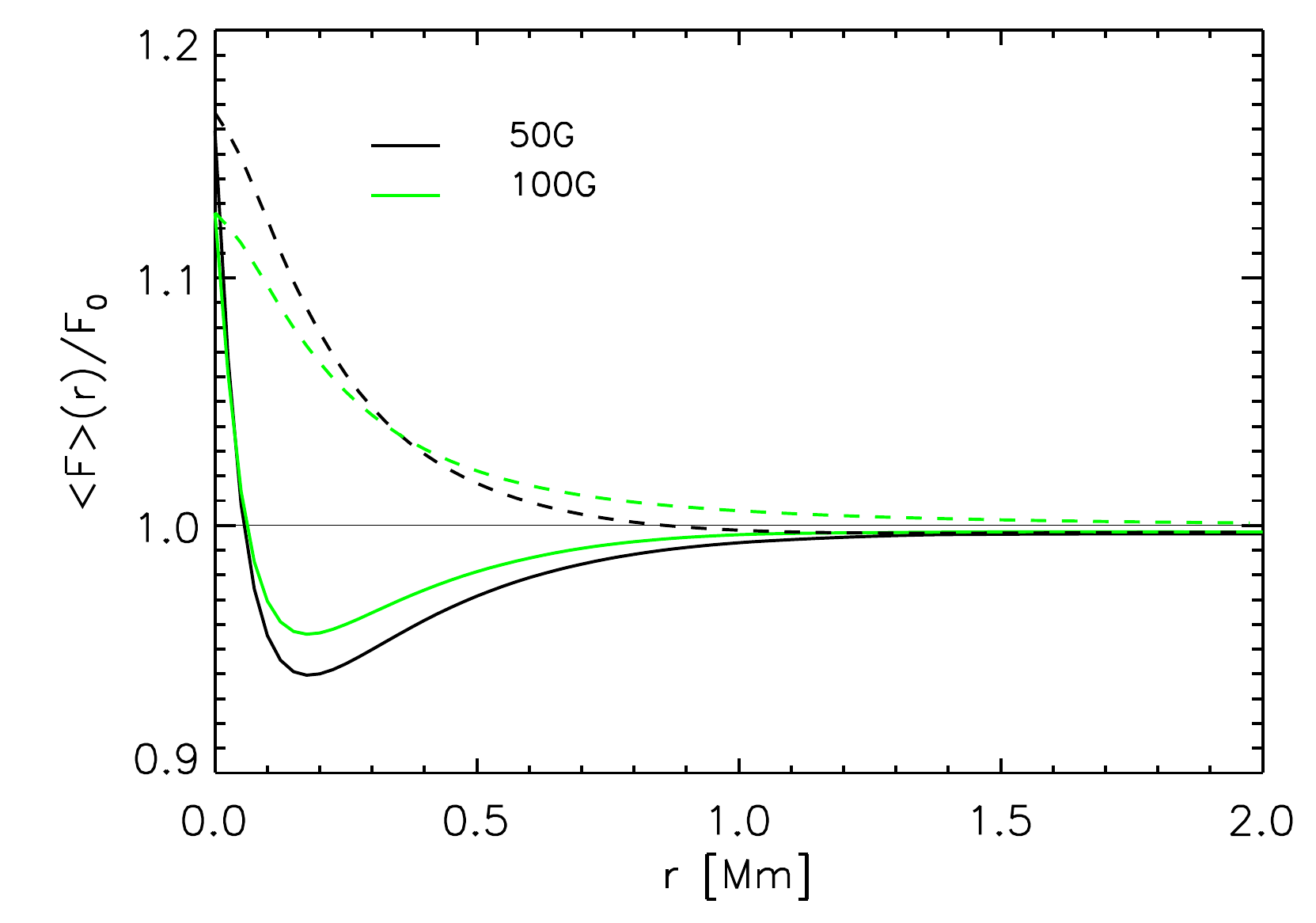}\hfill
\caption{Top,  solid lines: variation of bolometric flux as a function of distance $r$  from a magnetic concentration, corrected for dark lane bias, and using the nonmagnetic neighborhood selection method (see text).  Dashed: same but using the image superposition method of Fig.\ \ref{ringims}. Bottom: same data, but showing the average brightness inside the distance $r$. Green:  100 G simulation, black: 50 G simulation.}
\label{ringplots}
\end{figure}

\begin{figure*}[t]
\includegraphics[height=0.23\textwidth]{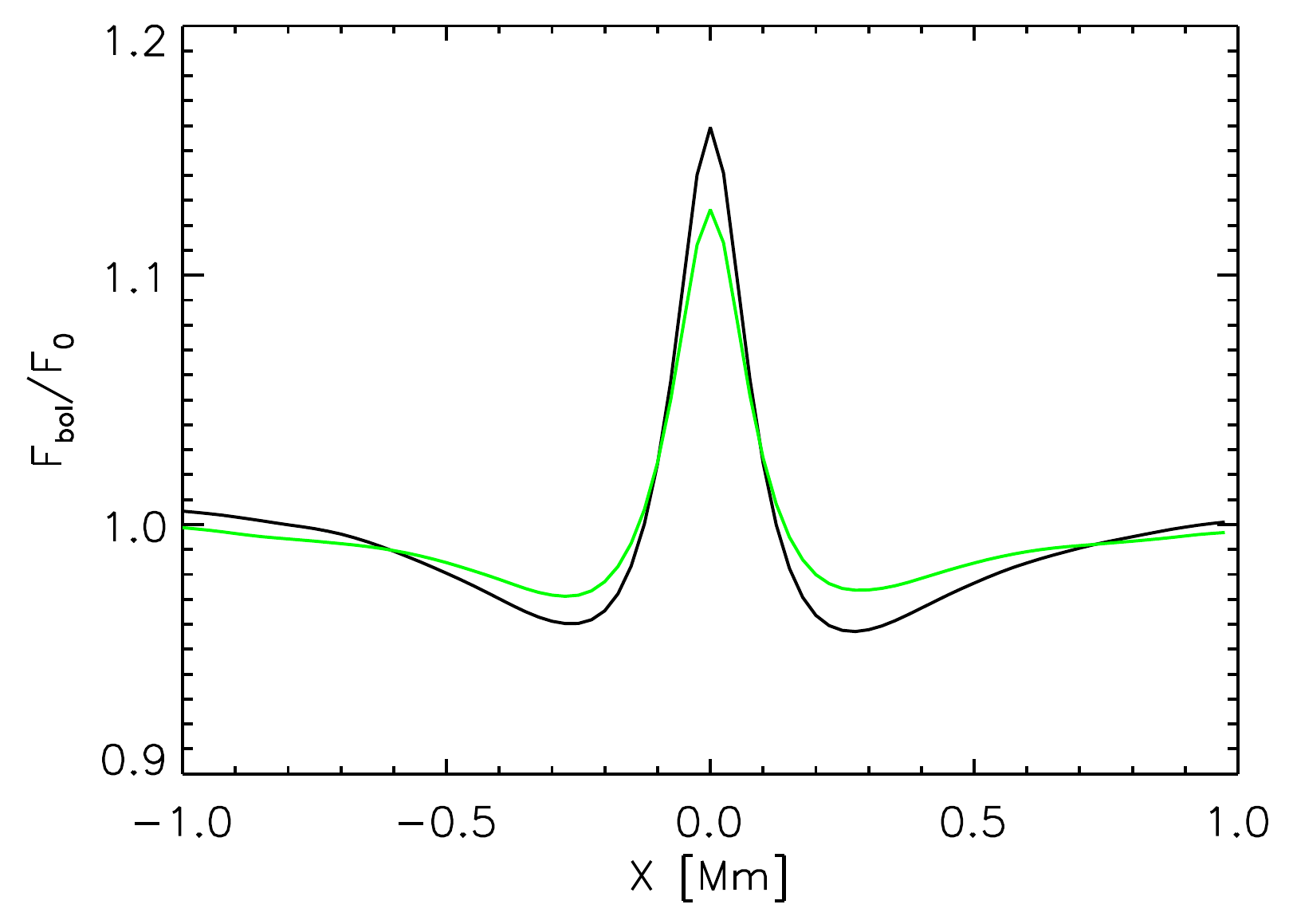}
\includegraphics[height =0.23\textwidth]{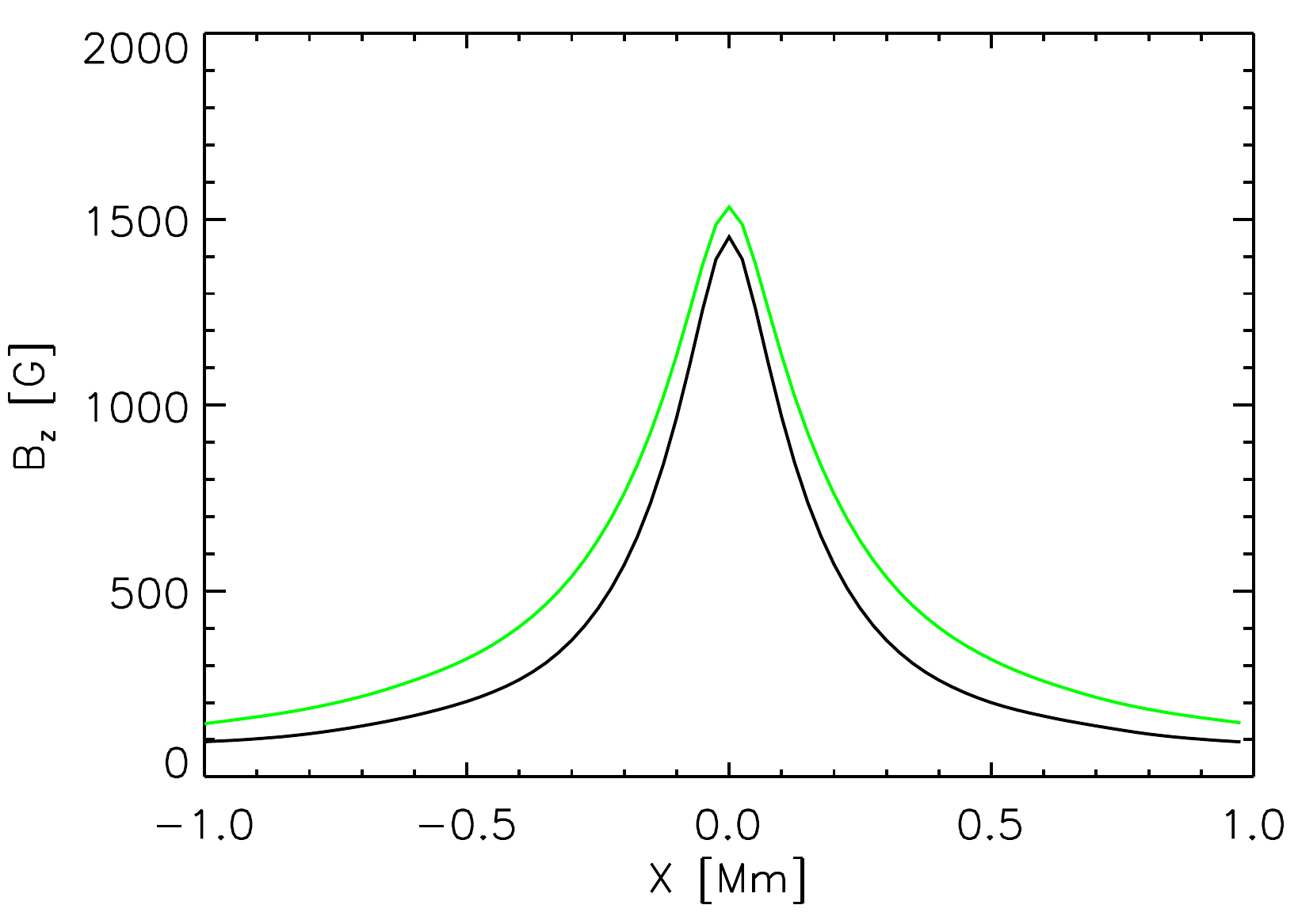}
\includegraphics[height =0.23\textwidth]{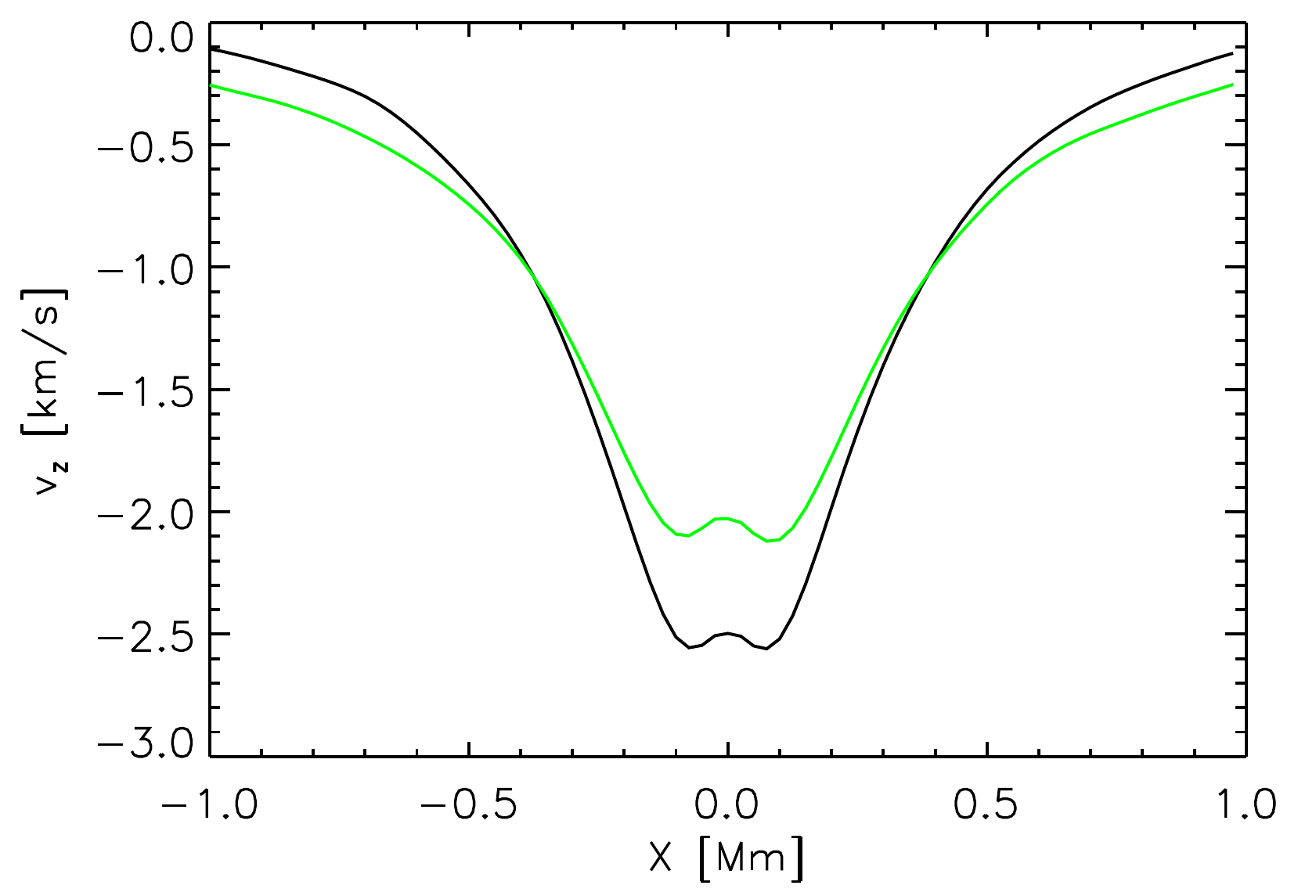}
\caption{Left: bolometric flux (normalized by the average flux $F_0$ of the nonmagnetic simulation). Middle: vertical magnetic field strength. Right: downflow velocity. All variables are shown along a straight line across the center of the averaged magnetic patch.  Green: 100 G simulation, black: 50 G.}\label{50-100}

\end{figure*}%

As a next step we choose representative nonmagnetic areas to which the environment  of the magnetic patch can be compared. We use the nonmagnetic simulation for this. The areas selected for comparison should  be as similar as possible to the ones where the field collects in the magnetic simulations.  
The typical environment magnetic structures sit in are the stagnation points of convective flows. As a practical definition of  stagnation points we choose those pixels where the convergence $\sigma=-{\rm div}\,{\bf v}_{\rm h}$ of the horizontal velocity ${\bf v}_{\rm h}$ is larger than  a minimum $\sigma_{\rm min}$. Its value is chosen such that the number of pixels selected is the same as the number selected in the magnetic image. This yields $\sigma_{\rm min}=0.049$s$^{-1}$ for the 50 G simulation. This is then our guess of the points where magnetic fields would concentrate if a magnetic field were present. For the 100 G simulation, where the  magnetic pixels occupy a larger area, a value  $\sigma_{\rm min}=0.039$s$^{-1}$ in the nonmagnetic images matches their larger number. 

A representative example of the convergence points selected in this way is shown in Fig.\ \ref{area_sel} (bottom panel). The average brightness image resulting from this selection is shown in  Fig.\ \ref{ringims} (middle panel). The center shows the darkening expected from an intergranular region. The brightness increases with distance from the stagnation point,  to a value near the average of the nonmagnetic surface. This image contains our estimate of the dark lane bias that is present in the magnetic image in the left panel. The difference between the two (right panel)  shows the magnetic brightness image corrected for the dark lane bias. It includes the sum of the proximity effect on nearby granulation and the dark rings. Due to the averaging over a large sample, the images are nearly axisymmetric. The remaining inhomogeneities give an impression of the noise level in the result. 

The dashed lines in Figure \ref{ringplots} (top panel) show the axisymmetric average of the right panel of Fig. \ref{ringims}, plotted as a function of distance from the centering pixel. The fluxes are normalized by the time- and area-averaged flux $F_0$ of the nonmagnetic simulation. In black are the results from the 50 G simulation, in green the 100 G simulation. The average brightness excess at the center of the magnetic patches is lower in the 100 G simulation than in 50 G simulation. This probably reflects the contribution of larger concentrations, whose properties start approaching those of pores. The intensity decrease due to the dark lanes is also less pronounced in the 100 G simulation. This leads to a lower integrated intensity effect of the magnetic patch environment in the case of the 100 G simulation compared to the 50 G simulation. A slightly different view is given in the left panel of Fig.\ \ref{50-100}, showing a 1-D section across the center of the averaged magnetic patch, comparing the 50 G and 100 G simulations.  The other two panels show the corresponding profiles of field strength and vertical velocity.

The middle panel of Fig.\ \ref{50-100} shows that the vertical magnetic field strength in the center of the magnetic patches is slightly higher in the case of the 100 G simulation than in the 50 G simulation. This difference increases with distance from the center of magnetic structure. As the average amount of magnetic flux in the concentrations is higher in the 100 G simulation than at 50 G, this is just an indication that the brightness per unit magnetic flux decreases somewhat with size. The main contribution to the intensity excess of the magnetic elements is the bright wall effect, which becomes conspicuous near the limb (Fig.\ \ref{compo}). Near disk center the positive contribution is mainly the brightening seen when looking down into the magnetic elements themselves (cf. S76). As the ratio of the perimeter to the area of the magnetic structures decreases with their size, the effect of the bright walls decreases as well. Because the dark ring effect is a direct consequence of the bright wall effect, it also decreases with increasing size of the magnetic structures. The combination leads to an smaller overall  intensity excess for magnetic structures of larger size, and explains the difference between the 100 G and the 50 G simulation.

\subsection{A more sensitive measurement}\label{sensitive_measurement}
Since the image superposition method described above produces a significant smearing of fine structure, an alternative procedure for quantifying the average brightness effects in the environment was devised. The selection of magnetic points on which to center the images is as before, but instead of superposing the entire surrounding image, the pixels used for constructing the average are restricted by the additional condition that they do not also satisfy the magnetic selection criterion. 

Call this the nonmagnetic neighborhood selection. With correction for the dark lane bias as before, the resulting variation of bolometric flux with distance is shown in Fig.\ \ref{ringplots} (top panel, solid lines). Compared with Fig.\ \ref{50-100} the environment of the magnetic point is now resolved much better. This is due to  the narrow elongated structure of magnetic concentrations. The additional selection emphasizes nearby pixels along the structure. The price is somewhat lower statistics, especially close to the center, but owing to the large area and time covered by the simulations the average remains well defined. Fig.\ \ref{ringplots}  (bottom panel) shows the effect on average brightness within a distance $r$ from the magnetic points. For large $r$, it approaches the mean brightness effect measured on the whole area of the simulation (-0.27\% and -0.34\% for the two simulations). It converges to this average roughly at a distance of 1.5 Mm.

\subsubsection{Dependence on patch size and  selection criteria}

The analysis method described in section \ref{sensitive_measurement} is insensitive to the fact that magnetic patches have different sizes. This is the case since the values at each radius, shown  as solid lines in Figure \ref{ringplots}, are averages over grid points which have the same distance from the last point qualifying as ``center of the magnetic patch'', irrespective of the size of the patch.  One could expect that the proximity effect of a large magnetic patch extends to a larger distance than that of a small one. This can be tested by subdividing the sample of patches into groups according to size. Because of limited statistics, two groups were used.

\begin{figure}[h]
\includegraphics[height=0.33\textwidth]{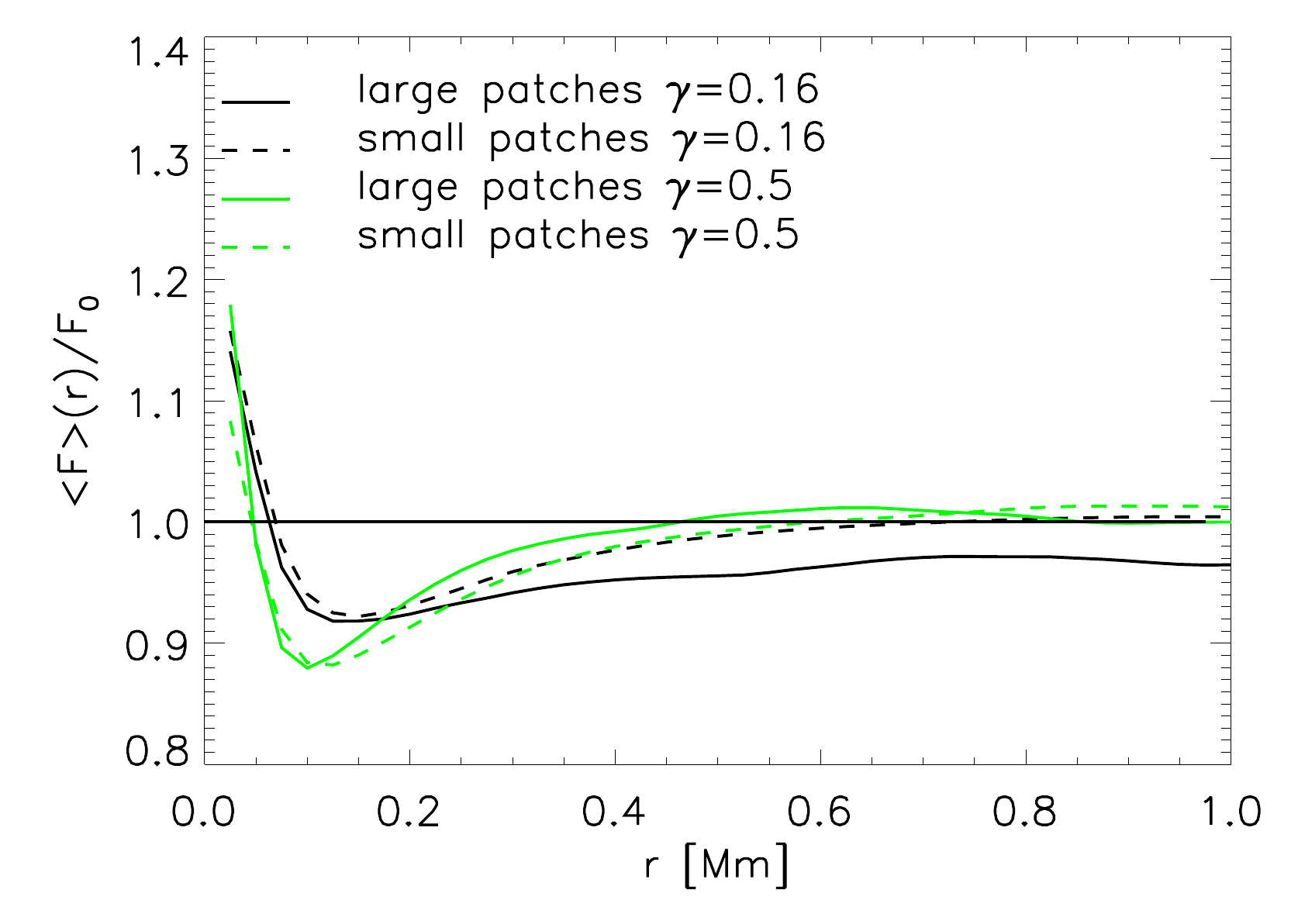}
\includegraphics[height=0.33\textwidth]{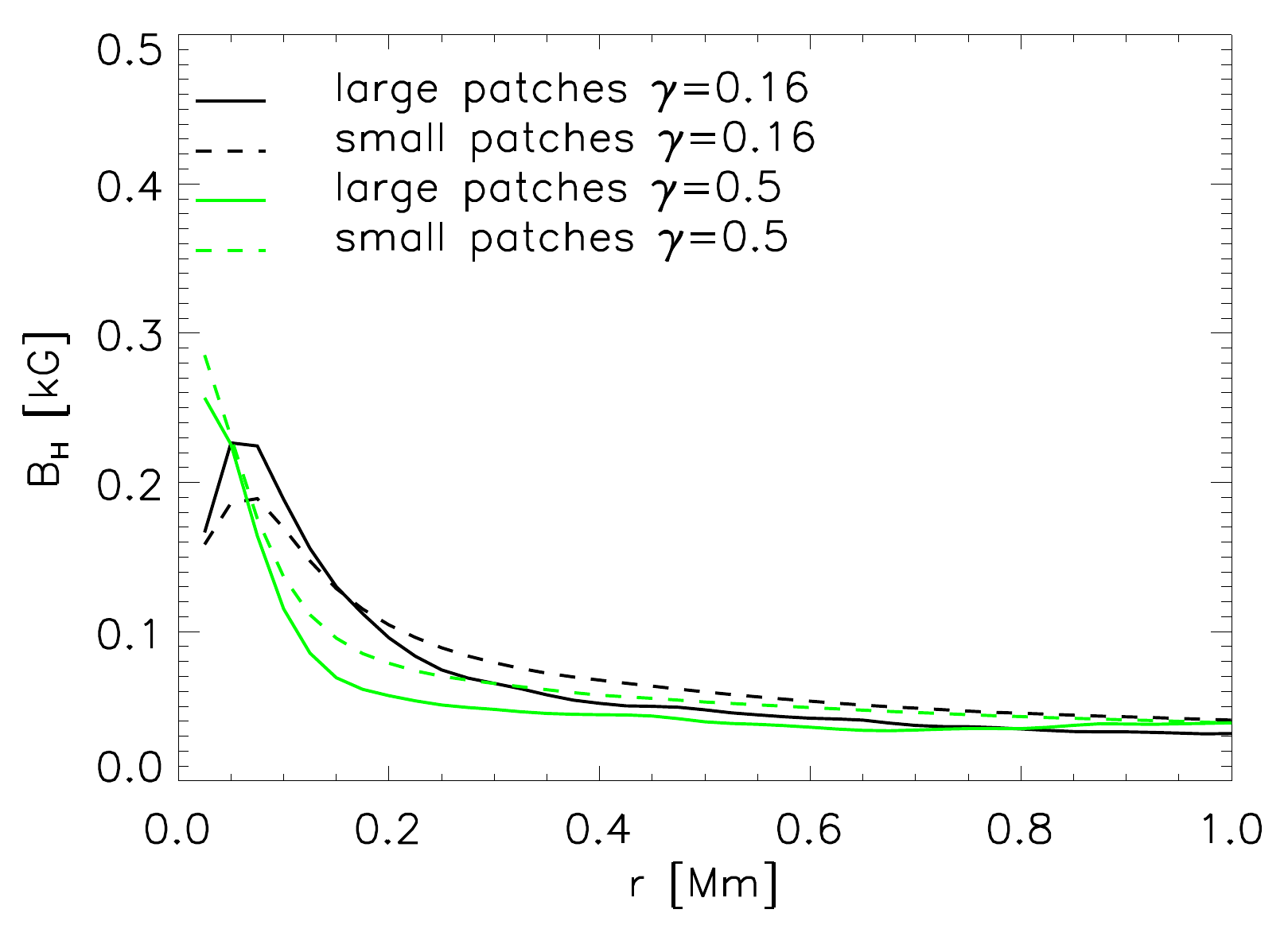}
\caption{Dependence on patch size and selection on field inclination. Top: bolometric flux, bottom: horizontal field strength 
 as a functions of distance for small (dashed lines) and large (solid) magnetic patches, for selection on two values  $\gamma=\vert B_{\rm h}/B_z\vert<$0.5 (green) or $\gamma<$1/6 (black). Selection on vertical magnetic field strength as before ($\vert B_z\vert> 1$ kG).}
\label{bol_int_gamma_dep}
\end{figure}

``Large" patches were defined as the largest 25\% of the magnetic patches in the sample, the rest as small ones. The  bolometric flux as a function of radius of the two groups is compared in Fig. \ref{bol_int_gamma_dep}. 
Close to the center of the magnetic patches the bolometric flux for small (dashed lines) and large (solid lines) patches looks very much the same, but at about 250-300 km away from the center they start to behave differently. The bolometric flux of the smaller patches slowly returns at this distance to the average bolometric flux value of the simulation, while the bolometric flux of the larger patches remains below the mean flux.  The influence of a magnetic element on the flow in its nonmagnetic environment thus appears to increase with its size.

To test how sensitive the result is to changes in criteria for inclusion of a pixel as belonging to a magnetic patch, we changed the condition on field line inclination. We required the field to be more vertical, a ratio of horizontal to total field strength $\gamma < $  1/6 instead of $\gamma< 0.5$.  The condition on vertical field strength was kept as before, $\vert B_z\vert> 1$ kG.
 The difference in the bolometric flux over radius between the two $\gamma$ values is shown in  Fig. \ref{bol_int_gamma_dep} (top). 
 To interpret this difference, the bottom panel of Fig. \ref{bol_int_gamma_dep}, shows the horizontal field strength as a function of distance from the center of the magnetic patches. From this figure it is evident that the $\gamma < $ 1/6  selects the center of the magnetic patches more strictly than $\gamma < 0.5$, which includes regions that are already at the outer edge of the magnetic patches. As a result the bolometric brightness excess at the center of the magnetic patches is higher for $\gamma <  $ 1/6. It also shows that the negative brightness contribution in the surroundings of the magnetic patches, a consequence of the dark ring effect and the convection suppression, is less pronounced for the stricter $\gamma$ selection. This is probably due to the fact that we are probing more nonmagnetic intergranular lanes in the case of the $\gamma < $ 0.5 selection. \\

Changing the selection criteria for ``large'' and ``small'' patches such that the subsample sizes are equal did not lead to significantly different results compared to the previous 25\% - 75\%  division. 
\subsubsection{Dependence on numerical resolution}
\begin{figure}[h]
\includegraphics[height=0.33\textwidth]{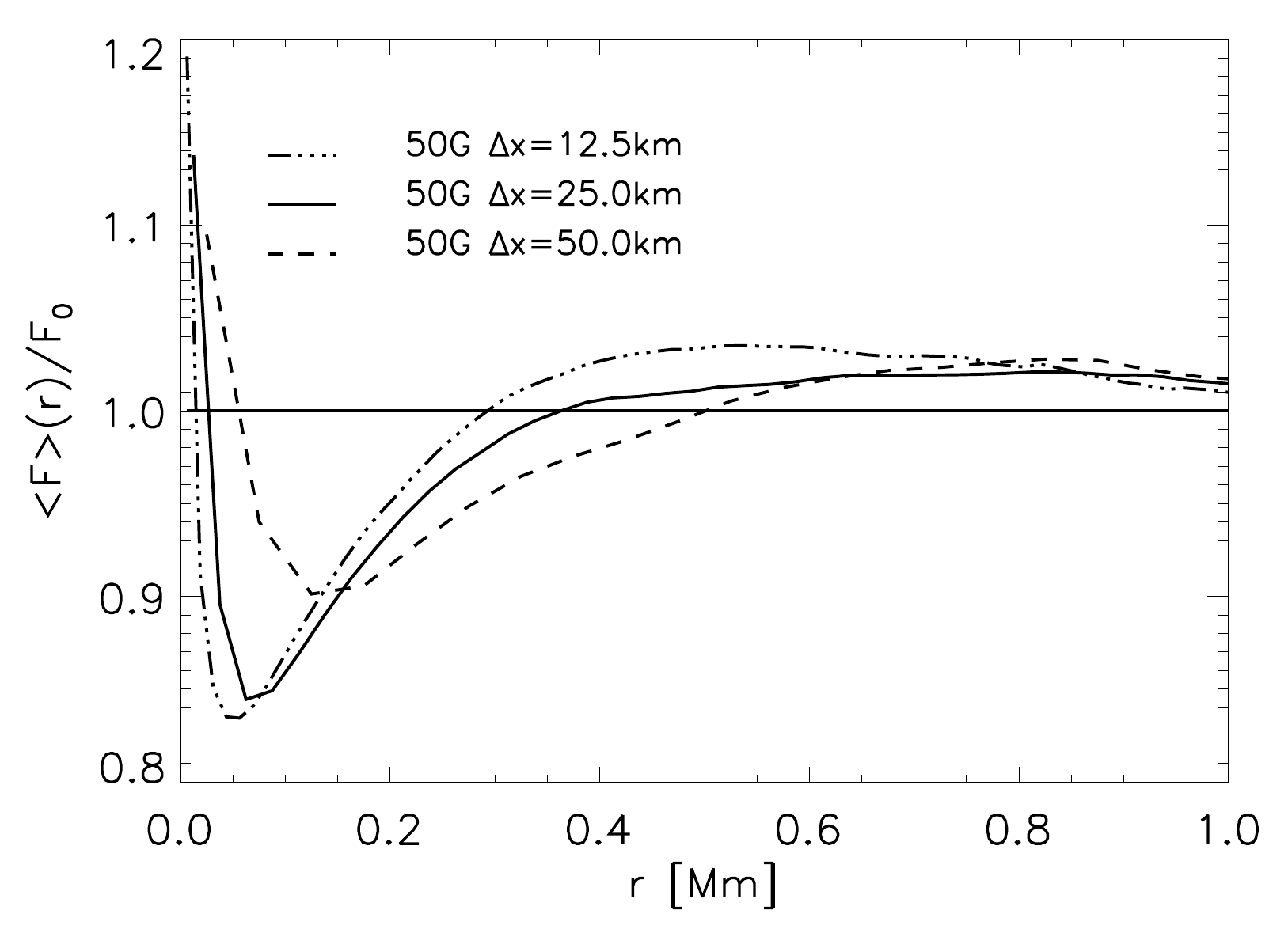}
\caption{Dependence  on numerical resolution. Bolometric flux as a function of distance for the 50 G simulation in a snapshot after 25 min simulation time. Horizontal resolution $\Delta$x=12.5 km (dot-dashed), 25 km (solid), 50 km (dashed).}
\label{bol_int_resolution}
\end{figure}

Fig \ref{bol_int_resolution} shows the bolometric flux over radial distance from the center of the magnetic elements for three different horizontal resolutions of $\Delta$x=12.5 km, 25 km and 50 km, for one snapshot after 25 min run time of a simulation 
with an initially uniform vertical magnetic field of 50 G.
 The change from 25 km resolution to 12.5 km mainly affects the brightness excess at the center of the magnetic elements, increasing it from 14\% to 19\% above the mean bolometric flux value, while the dark ring effect is amplified by only 1\%. Switching from 50 km resolution to $\Delta$x=25 km affects the brightness contributions from the dark ring region and the center of the magnetic elements almost equally; these effects are amplified by around 5\% in the higher resolution. But one has to be careful comparing the bolometric flux over radius for different numerical resolutions like this. Even though the same initial snapshot was used for the different spacial resolutions, a different spatial resolution also leads to different evolution of the granulation. The change of the granulation pattern introduces a statistical fluctuation of the bolometric flux which can not be disentangled from the numerical resolution effect, unless one has long enough runs times to quantify the statistical fluctuations. This requires a computational effort beyond the scope of this work.
\\ 

\subsection{Vertical velocities}
The dark ring occurs because of radiative cooling in the surroundings of the magnetic elements. This cooling means that enhanced downdrafts are expected. The effect is seen in the righthand panel of Fig.\ \ref{50-100}, though the very narrow structure  of the downdrafts has been smeared out considerably by averaging process. The average downflow speeds are consequently also less pronounced in the 100 G simulation than in the 50 G simulation.

As was done in Fig.\ \ref{ringplots} for the bolometric flux, the effect of the magnetic fields on their surroundings can also be seen in the  vertical velocity amplitudes near magnetic concentrations. We compare them with the velocities in similar regions in a nonmagnetic simulation. They are selected on the basis of flow convergence (the locations where the small scale magnetic field is expected to collect), with the same selection process as used for the bolometric flux difference.  The result (Fig.\ \ref{velocity}) shows that the downflows are stronger around the magnetic elements, as expected. With increasing distance the average velocity becomes dominated by  upflows in the surrounding granulation. Up to a distance of about 700 km, the upflow speeds are markedly lower around magnetic elements. Beyond this distance the sign of the difference reverses. The velocity difference (bottom panel) peaks around $-1.1$ km/s, at a distance of $\approx 300$ km.  Note the similarity of these difference curves to the bolometric flux differences in Fig. \ref{ringplots} (top).

\section{Discussion and conclusions}
The effect of the small scale magnetic field on (bolometric) brightness appears to have three distinct components: the brightness of the magnetic structure itself  (composed of the bright interior of the structure at disk center, and the`bright wall effect'  towards the limb), plus the two proximity effects it has on its surroundings: the `dark ring' resulting from the influx of radiation into the magnetic concentration, and the interference of magnetic concentrations with the nearby convective flow. 

The most conspicuous component is the bright wall effect, easily measurable as a brightening in active regions when seen near the solar limb. It has also been reproduced convincingly in realistic 3-D MHD simulations such as Carlsson et al.\ (2004), De Pontieu et al.\ 2006, and the present ones (cf.\ Fig.\ \ref{compo}). The $0\farcs1-0\farcs2$ narrow dark rings are also conspicuous in high resolution continuum images near disk center, but are less easily quantifiable because of the variable shapes of the structures (`crinkles'). Finally, the effect on the surrounding convective flow pattern is well known from observations, but its effect on brightness is hard to detect, smeared out over too large an area to be measurable at the photometric accuracy of ground-based observations. It has been detected however, in data from Hinode (Kobel et al. 2012).

\begin{figure}[t]
\center\includegraphics[width=0.9\hsize]{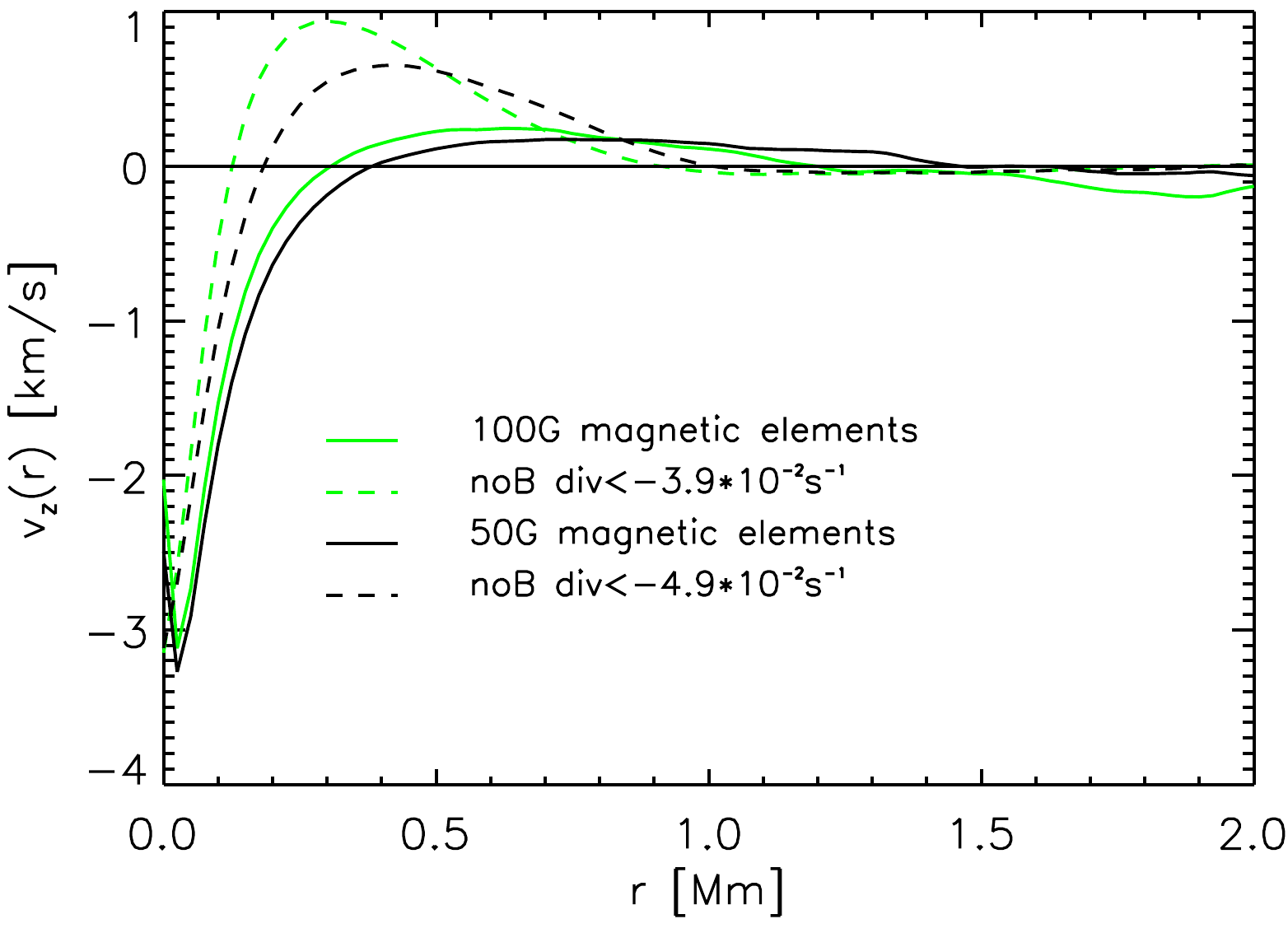}
\center\includegraphics[width=0.9\hsize]{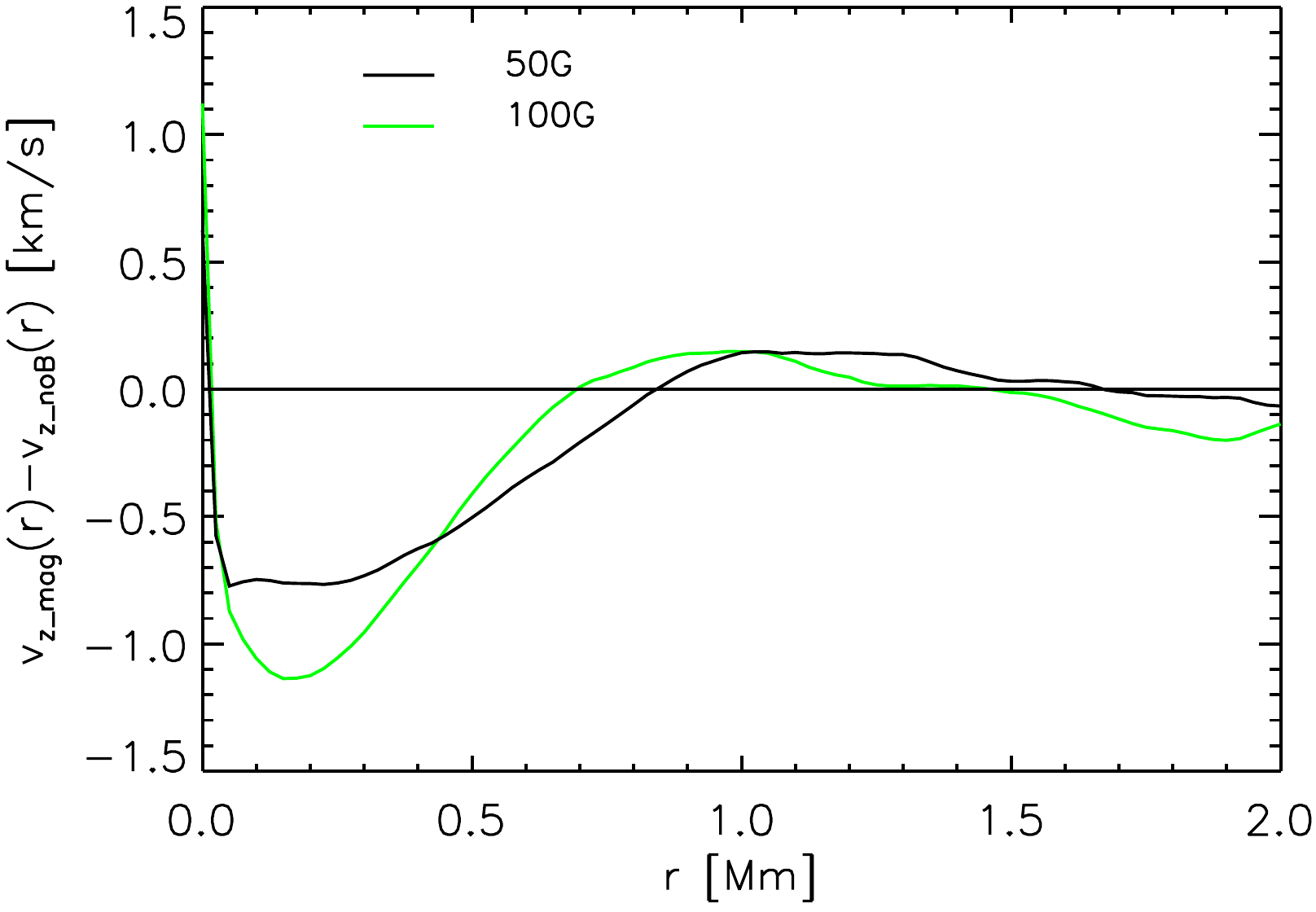}
\caption{Top: average vertical velocity  $\langle v_z\rangle$ as a function of distance from magnetic elements (solid) compared with  $\langle v_z\rangle$ near flow convergence points of the nonmagnetic simulation (broken). Bottom: difference between the two, showing the proximity effects on vertical velocities.}\label{velocity}
\end{figure}

Perhaps surprisingly, the negative contribution of the proximity effects  turns out to dominate the photospheric brightness change. The net brightness effect is thus the opposite of the  standard prediction (S77). 
Observationally, the effect of magnetic concentrations on nearby convective flows is easily detectable through changes in granulation morphology and vertical velocities (`anomalous granulation', cf. Title et al. 1986, 1992, Kobel et al. 2012). Concerns that these changes could also affect energy transport and hence surface flux have been around for some time (e.g.,\ Spruit 1998). They were not discussed much, possibly because the effect was not large enough to be detected with ground-based photometric accuracy. Our simulations also show changes in vertical velocities near magnetic concentrations. The spatial coincidence of these changes with changes in bolometric flux support the interpretation that the net darkening is caused by interference with the convective heat flux near magnetic concentrations. The effect appears to take place in a rather narrow region, extending from the intergranular lane to somewhat into the surrounding granule. 

The magnetic concentrations are larger on average in the 100 G simulation than at 50 G. Their effect on the surrounding flows is correspondingly somewhat larger (Fig.\ \ref{velocity} bottom panel). The net negative effect on bolometric brightness does not differ much in the 50 G simulation. We interpret this as a consequence of the compensating bright wall effect. Its increase towards the limb is most prominent in larger concentrations that are less affected by self-obscuration away from disk center. This is evident in the disproportionately larger brightness increase of the 100 G simulation towards the limb compared with the 50 G result (Fig.\ \ref{clv}). It can also be seen qualitatively in the CLV of the images in Fig.\ \ref{compo}.   

This leaves the question how the observed positive correlation of total solar irradiance (TSI) with the small scale field comes about. The most likely explanation is that the simulations underestimate the contribution of chromosphere and upper photosphere. A major contribution of the chromosphere to TSI  has in fact already been inferred from the wavelength dependence of solar irradiance variability. Unruh et al. (1999) concluded that the photosphere contributes negligibly to TSI variation, compared with the chromosphere. More recently, Ball et al. (2012) estimate the photospheric contribution at 18\%.

Empirical models for the mean stratification in active regions such as Vernazza et al. (1973) already indicated the presence of heating processes starting around the temperature minimum. This has been interpreted as evidence of some form of magnetic heating. Our calculations necessarily miss most of this contribution because of the use of a potential field as upper boundary condition. This forces the field near the upper boundary to its lowest energy state, from which no energy can be extracted. Proper inclusion of magnetic dissipation in simulations like the present ones is very demanding, because of the time step limitations resulting from the high Alfv\'en speed in the chromosphere. Simulations with methods adapted to this situation may be needed, such as have been developed by Gudiksen and Nordlund (2005) for the coronal heating problem. \\

\acknowledgements{Based in part on data from the Swedish 1-m Solar Telescope, operated on the island of La Palma by the Institute for Solar Physics of the Royal Swedish Academy of Sciences in the Spanish Observatorio del Roque de los Muchachos of the Instituto de Astrof\'{\i}sica de Canarias}.

\end{document}